\documentclass[journal]{IEEEtran}
\usepackage{amssymb}
\usepackage{makecell}
\usepackage{multirow}
\usepackage{graphicx}
\usepackage{subcaption}
\usepackage{graphicx}
\usepackage{cite}
\usepackage{bm}
\usepackage{pifont}
\usepackage{bbding}
\usepackage{amsmath}
\usepackage{amssymb}
\usepackage{mathrsfs,balance}
\usepackage{algorithm}
\usepackage{algorithmic}
\usepackage{color}
\usepackage{amsthm}
\usepackage{booktabs}
\usepackage{stfloats}
\usepackage{xcolor}
\usepackage{amsthm}

\usepackage{amsthm}
\usepackage{amssymb}

\newtheorem{proposition}{Proposition}

\begin{document}
\title{Asynchronous Uplink Multi-UE Networked Sensing via Multipath Correlation}

\author{
        Jingying~Bao,
        J. Andrew~Zhang,~\IEEEmembership{Senior Member,~IEEE},
        Kai~Wu,~\IEEEmembership{Member,~IEEE},
        Christos~Masouros,~\IEEEmembership{Fellow,~IEEE},
        and~Y. Jay~Guo,~\IEEEmembership{Fellow,~IEEE}

\thanks{J. Y. Bao, J. A. Zhang, K. Wu and Y. J. Guo are with the Global Big Data Technologies Centre, University of Technology Sydney, Sydney, NSW 2007, Australia (e-mail: Jingying.Bao@student.uts.edu.au; andrew.zhang@uts.edu.au; kai.wu@uts.edu.au; jay.guo@uts.edu.au).}
\thanks{C. Masouros is with the Department of Electronic and Electrical
Engineering, University College London, London WC1E 7JE, U.K. (e-mail:
c.masouros@ucl.ac.uk)}
}
\maketitle
\begin{abstract}
Multiuser uplink sensing offers a promising paradigm for networked sensing by leveraging signals from multiple user equipment (UEs) to enhance sensing coverage and performance. This paper investigates asynchronous multiuser uplink orthogonal frequency-division multiple access sensing by exploiting correlated multipath structures across UEs. A common frontend first mitigates the dominant timing offsets (TOs), carrier-frequency offsets (CFOs), and phase offsets (POs) through LoS-referencing and constructs a reduced delay-angle-of-arrival (AoA) search space. Two alternative multiuser sensing estimators are then proposed, with different mechanisms of explicitly and implicitly exploiting UE correlation, respectively. The first explicitly adopts a latent-group shared-private sparse Bayesian learning (Group-SBL) model to exploit group-shared propagation structures while accommodating UE-specific components. The second combines calibrated multiuser observations for joint multiple signal classification (MUSIC) processing, recovering a global delay-AoA pool that can be further associated with individual UEs. Residual-TO refinement is conducted with both estimator updates, followed by local-grid delay-AoA refinement and residual common-phase compensation before Doppler estimation. Simulations under a 5G New Radio-compatible uplink configuration show that both multiuser schemes outperform their corresponding per-UE baselines, with larger gains as cross-UE propagation correlation increases.
\end{abstract}
\begin{IEEEkeywords}
Networked sensing, perceptive mobile network (PMN), multi-user OFDMA, sparse Bayesian learning (SBL), clock asynchronism.
\end{IEEEkeywords}
\section{Introduction}
Integrated sensing and communication (ISAC) is increasingly viewed as a key capability of future wireless networks, where existing communication infrastructures are endowed with environmental sensing functions rather than serving solely as data-delivery platforms \cite{ref_isac_jsac}. This evolution naturally leads to perceptive and networked sensing architectures, in which sensing information acquired at distributed network nodes can be processed cooperatively or centrally to improve environmental perception \cite{ref_network_level_isac}. Among such architectures, uplink sensing is particularly attractive
because it reuses existing user equipment (UE) transmissions
at the base station (BS) for environmental sensing~\cite{ref_pmn_framework},
avoiding dedicated sensing transmitters and reducing
hardware and deployment complexity~\cite{ref_pmn_survey}. 
More importantly, UEs observing similar environments may provide correlated and complementary observations of common targets or scattering regions. Aggregating such multiuser information can improve sensing reliability through cooperative and spatial diversity \cite{ref_networked_isac}.

Realizing these potential gains in multiuser uplink sensing, however, involves two major challenges. The first is to mitigate synchronization errors caused by independent UE oscillators and clocks, which introduce timing offsets (TOs), carrier-frequency offsets (CFOs), and phase offsets (POs). These offsets impair coherent sensing processing \cite{ref_bistatic_spm,ref_jump}.  The second is how to exploit the underlying correlation structure across UEs, particularly considering the partially shared multipath structure, i.e., as different UEs may observe both common and UE-specific multipath components with variations in their propagation parameters \cite{ref_dai_grouping}.

Synchronization impairments have been extensively studied in wireless sensing, with representative solutions including cross-antenna channel state information (CSI) ratio processing \cite{ref_witraj} and reference- or parameter-estimation-based compensation \cite{ref_jump,ref_efficient_async}. Residual synchronization errors after over-the-air synchronization were further analyzed in bistatic orthogonal frequency-division multiplexing (OFDM) sensing \cite{ref_bistatic_async}, while \cite{ref_pilot_sfo} improved residual sampling-frequency offset estimation to suppress range-Doppler migration. More recently, \cite{ref_subspace_async_residual} incorporated residual-TO compensation into subspace-based angle-delay estimation. However, a scheme that accommodates distinct offsets across multiple UEs while preserving the correlation structure for subsequent processing has yet to be developed.

Structural sparsity and correlation have been extensively
exploited in communication channel estimation, including
spatially common sparsity \cite{ref_common_sparsity}
and group-sparse Bayesian channel estimation
\cite{ref_group_sbl_mmwave}. Beyond fixed correlation structures, \cite{ref_dai_grouping} jointly performed channel estimation and user grouping by decomposing each channel into shared and individual sparse components. In \cite{ref_dp_cluster}, Bayesian clustering was used to group reconfigurable intelligent surface (RIS) sub-surfaces with similar sparse channel structures and exploit their common sparsity. These studies primarily target communication channel or CSI reconstruction, rather than sensing parameter estimation. Shared-private sparsity has also been exploited to capture
cross-UE delay correlation in asynchronous multiuser
uplink sensing \cite{ref_prior_tvt}.
Meanwhile, distributed and cooperative multiple signal classification (MUSIC) methods have also been developed to jointly exploit observations from multiple sensing nodes. For example, \cite{ref_geometry_music2} fused local two-dimensional MUSIC spectra across distributed radars for joint range-angle estimation, while \cite{ref_multistatic_music} combined multistatic MUSIC spectra for multi-target localization. Recently, \cite{ref_joint_multistation_music} employed a unified multi-platform covariance model for joint angle estimation and target association. Although these methods demonstrate the benefits of multi-node subspace processing, they primarily exploit geometric consistency among observations of common targets, without explicitly leveraging cross-UE correlations arising from partially shared multipath structures.

To address these challenges, we develop a multiuser uplink sensing
framework combining synchronization offset mitigation with cross-UE
delay-angle-of-arrival (AoA) correlation exploitation.
A common frontend mitigates dominant synchronization offsets
and constructs a reduced delay-AoA search space.
Two alternative sensing estimators, based on Bayesian sparse modeling
and joint MUSIC (JMUSIC) processing, exploit cross-UE parameter
correlation in distinct ways while accommodating UE-dependent
frequency-domain observations under orthogonal
frequency-division multiple access (OFDMA) resource allocation.
The main contributions are as follows:

\begin{itemize}

\item We develop an offset-mitigation and sensing framework for asynchronous multiuser uplink sensing. A LoS-referenced frontend mitigates the TO, CFO, and PO components, while residual-TO refinement is conducted with the delay-AoA estimator updates. After delay-AoA refinement, the residual common phase is further compensated before Doppler estimation, improving robustness to imperfect synchronization.

\item We propose a latent-group shared-private sparse Bayesian learning
(Group-SBL) scheme that exploits cross-UE delay-AoA correlation
through group-level shared sparsity. The variational inference jointly learns the latent UE grouping and shared-private sparse structures, thereby explicitly exploiting UE correlation through commonly supported components while estimating UE-specific paths.

\item We also develop a JMUSIC scheme that implicitly exploits cross-UE information through multiuser subspace aggregation without explicit probabilistic correlation modeling. The calibrated multiuser observations are combined to recover a global delay-AoA pool, from which UE-specific propagation supports are obtained through concentrated deterministic maximum-likelihood (DML) association.

\end{itemize}

Simulation results under a 5G New Radio (NR)-compatible uplink configuration validate the effectiveness of the proposed framework. Both Group-SBL and JMUSIC outperform their corresponding per-UE baselines in delay-AoA estimation, demonstrating the benefits of exploiting cross-UE propagation correlation. 
The results show that refining residual synchronization errors improves sensing robustness under imperfect calibration.
 Further simulations reveal the tradeoff between multiuser correlation gains and per-UE sensing resources. 

The remainder of this paper is organized as follows.
Section~II presents the system model.
Section~III introduces the common signal processing procedures,
including offset mitigation, reduced candidate construction,
and residual synchronization error modeling.
Sections~IV and V present the Group-SBL and JMUSIC schemes,
respectively, with their corresponding residual-TO refinement procedures.
Section~VI describes the common back-end processing
for local-grid delay-AoA refinement,
residual common-phase compensation, and Doppler estimation.
Section VII analyzes the computational complexity,
Section VIII presents the simulation results,
and Section IX concludes the paper.
\section{System Model}
\label{sec:system_model}

\begin{figure}[t]
    \centering
    \includegraphics[width=0.48\textwidth]{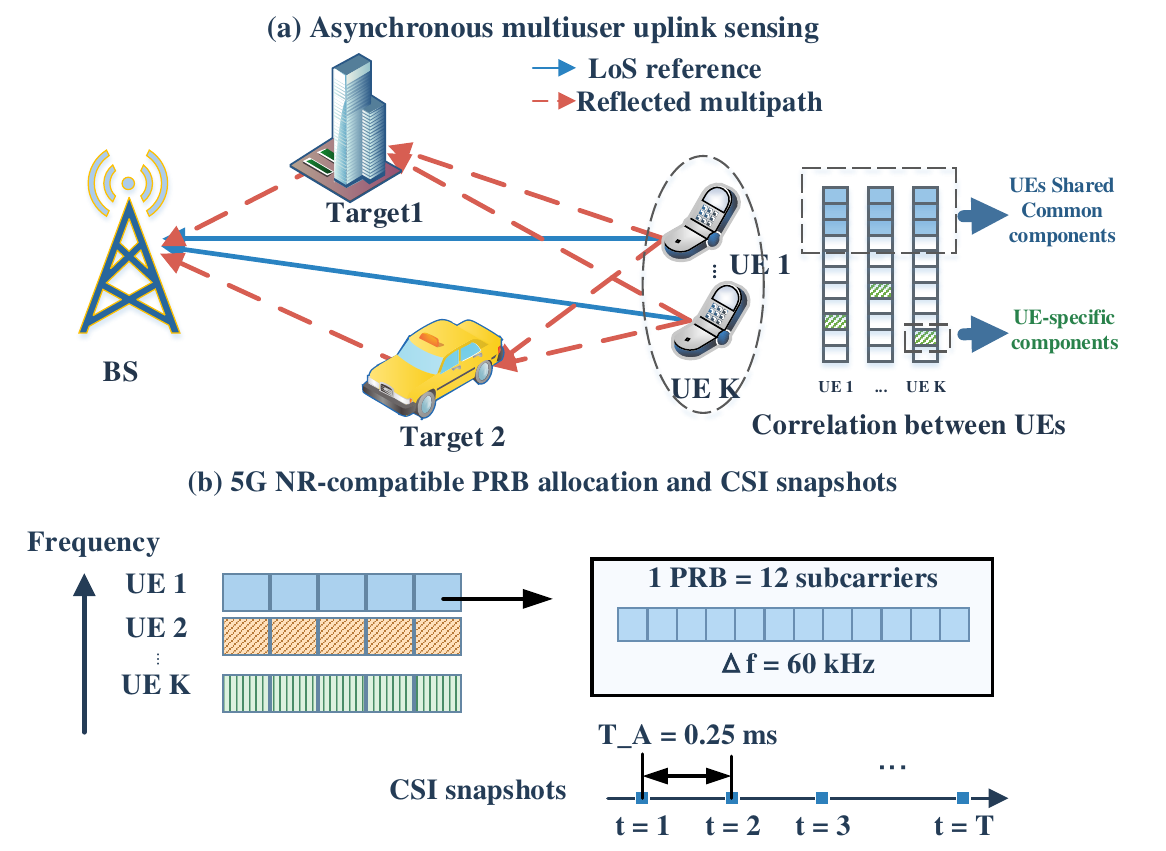}
    \caption{Multiuser uplink OFDMA sensing scenario.}
    \label{fig:system_model}
\end{figure}

In Fig.~\ref{fig:system_model}, we consider a multiuser uplink sensing scenario where a BS equipped with an
$M$-element uniform linear array (ULA) receives OFDMA signals from $K$
single-antenna UEs. The active subcarriers, spaced by $\Delta f$, are organized
into physical resource blocks (PRBs), each containing 12 consecutive
subcarriers. Following localized physical uplink shared channel (PUSCH) resource allocation, UE $k$ is assigned
a contiguous set of PRBs, with subcarrier-index set
$\mathcal{N}_k$ with $N_k=|\mathcal{N}_k|$ and
$\mathcal{N}_k\cap\mathcal{N}_{k'}=\varnothing$ for $k\neq k'$.
A CSI observation is extracted at each of $T$ snapshots at
$t_t=(t-1)T_A$, where $T_A$ denotes the snapshot interval. CSI observations are assumed available
over all allocated subcarriers, without explicitly
modeling the pilot resource element mapping.

Let $L_k$ denote the number of resolvable propagation paths associated with
UE $k$. The $\ell$th path is characterized by the complex gain
$\alpha_{k,\ell}$, propagation delay $\tau_{k,\ell}$, Doppler frequency
$\nu_{k,\ell}$, and AoA $\theta_{k,\ell}$. For the ULA,
the spatial response is
$\mathbf a(\theta)
=\frac{1}{\sqrt M}
[1,e^{-j2\pi d\sin\theta},\ldots,
e^{-j2\pi d(M-1)\sin\theta}]^{\mathsf T}$,
where $d$ is the inter-element spacing normalized by the wavelength.
Over the subcarriers assigned to UE $k$, the delay response is
$\mathbf{d}_k(\tau)
=[e^{-j2\pi f_{k,1}\tau},\ldots,
e^{-j2\pi f_{k,N_k}\tau}]^{\mathsf T}$,
where $f_{k,n}$ is the baseband frequency of the $n$th allocated subcarrier.

Due to the lack of synchronization between the UEs and the BS \cite{ref_jump,ref_bistatic_async}, the CSI
observations are subject to snapshot-dependent TO, CFO, and PO, denoted by
$\delta\tau_k[t]$, $\delta f_k[t]$, and $\beta_k[t]$, respectively.
After removing the known transmitted symbols, the resulting CSI matrix of
UE $k$ at the $t$th snapshot is modeled as
\begin{equation}
\begin{aligned}
\mathbf{Y}_k[t]
={}&
e^{j\beta_k[t]}
\sum_{\ell=1}^{L_k}
\alpha_{k,\ell}
e^{j2\pi\left(\nu_{k,\ell}+\delta f_k[t]\right)t_t}
\\[-1mm]
&\quad\times
\mathbf{a}(\theta_{k,\ell})
\mathbf{d}_k^{\mathsf T}
\!\left(\tau_{k,\ell}+\delta\tau_k[t]\right)
+\mathbf{E}_k[t].
\end{aligned}
\label{eq:async_received_signal}
\end{equation}
where $\mathbf{E}_k[t]$ denotes the additive noise.

As suggested in \cite{ref_dai_grouping}, nearby UEs, as shown in
Fig.~\ref{fig:system_model}, may exhibit partially correlated
propagation structures due to similar propagation environments.
Specifically, different UEs may observe shared multipath
components with similar delay-AoA parameters, while also
having non-shared UE-specific components. This partial overlap in
multipath structures gives rise to cross-UE correlation
in their sensing parameters. UEs exhibiting similar shared
multipath structures can thus be grouped together. Motivated by this observation,
we develop two alternative multiuser sensing estimators
to exploit such correlations while accommodating
UE-specific multipath components.

%Practical sensing objects may give rise to multiple closely spaced propagation paths due to their spatial extent and surrounding scatterers. Therefore, nearby UEs observing the same target or scattering region may exhibit slightly different delays and AoAs. We accordingly model the multiuser propagation structure using correlated delay-AoA supports rather than requiring them to be strictly identical.

%In this work, the framework is instantiated by jointly estimating correlated delay-AoA structures across UEs, which characterize the frequency-spatial signature of each path. 

\section{Common Signal Processing}

\begin{figure}[t]
    \centering
    \includegraphics[width=0.5\textwidth]{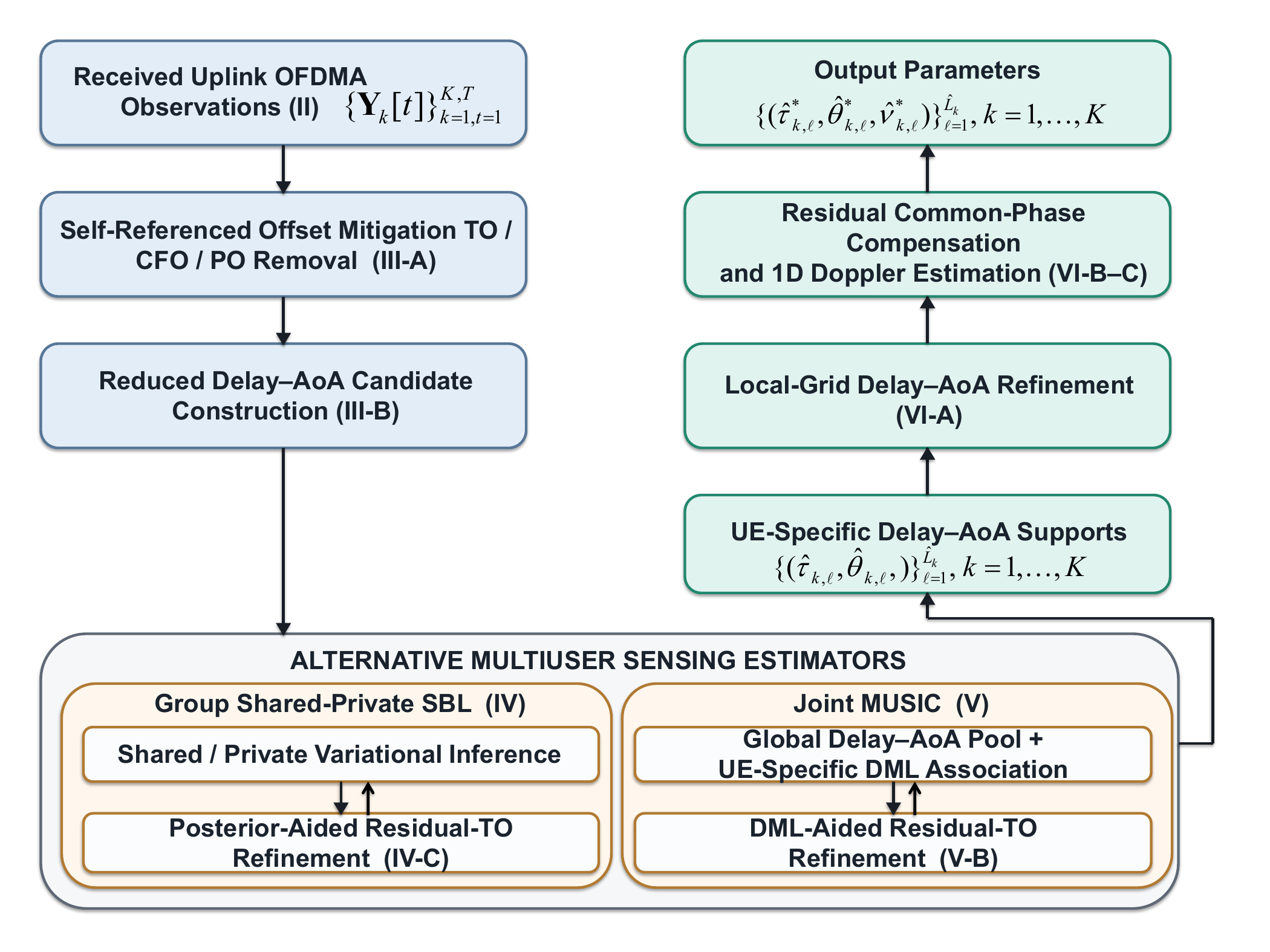}
\caption{Proposed multiuser uplink sensing framework.}
    \label{fig:overall_framework}
\end{figure}

The overall processing flow of the proposed multiuser sensing framework
is illustrated in Fig.~\ref{fig:overall_framework}. The received uplink
OFDMA observations are first processed by a common front end, which
performs LoS-referenced offset mitigation
using the dominant LoS path as a reference
and constructs a reduced delay-AoA candidate set. Two alternative multiuser sensing estimators
then operate on the reduced candidate set: the Group-SBL scheme presented in Section~IV and the JMUSIC scheme in Section~V. In both branches, residual-TO refinement is conducted with the respective estimator updates. After convergence, the resulting UE-specific delay-AoA supports are processed by the common back end in Section~VI, comprising local-grid delay-AoA refinement, residual common-phase compensation, and one-dimensional Doppler estimation.

\subsection{Self-Referenced Offset Mitigation}

The offsets in \eqref{eq:async_received_signal} shift the delay responses and
rotate the snapshot phases, thereby obscuring the cross-UE delay-AoA
correlation. Since these offsets are common to all paths of a given UE within
each snapshot, the dominant quasi-static line-of-sight (LoS) component is used as a reference.
For the LoS reference path, its observed delay is
$\rho_k[t]\triangleq\tau_{k,\mathrm{ref}}+\delta\tau_k[t]$, and the common
snapshot phase is
$\psi_k[t]\triangleq2\pi\delta f_k[t]t_t+\beta_k[t]$.

To improve the reliability of LoS reference estimation, we jointly exploit the spatial and frequency-domain information through a two-dimensional delay-AoA formulation. By separating the search into successive one-dimensional AoA and delay estimations, the proposed procedure avoids the computational burden of an exhaustive two-dimensional search. 

As the spatial power metric is insensitive
to synchronization offsets, the reference AoA is first searched over
$G_{\theta}^{\mathrm{ref}}$ uniformly spaced grid points
$\{\theta_q^{\mathrm{ref}}\}_{q=1}^{G_{\theta}^{\mathrm{ref}}}$ on
$[-\pi/2,\pi/2]$ using the cross-snapshot spatial score
$\frac{1}{T}\sum_{t=1}^{T}
\|\mathbf a^{\mathsf H}(\theta_q^{\mathrm{ref}})
\mathbf Y_k[t]\|_2^2/
\|\mathbf a(\theta_q^{\mathrm{ref}})\|_2^4$,
whose $C_{\mathrm{ref}}$ dominant peaks are retained as the coarse AoA
candidates
$\{\widehat{\theta}_{k,c}^{\mathrm c}\}_{c=1}^{C_{\mathrm{ref}}}$.

For candidate $c$, the observed delay is estimated by spatial
projection over $G_{\rho}^{\mathrm{ref}}$ uniformly spaced grid points
$\{\rho_g^{\mathrm{ref}}\}_{g=1}^{G_{\rho}^{\mathrm{ref}}}$ on
$[0,\tau_{\max}]$ as
\begin{equation}
\widehat{\rho}_{k,c}[t]
=
\arg\max_{\rho\in
\{\rho_g^{\mathrm{ref}}\}_{g=1}^{G_{\rho}^{\mathrm{ref}}}}
\left|
\frac{
\mathbf a^{\mathsf H}(\widehat{\theta}_{k,c}^{\mathrm c})
\mathbf Y_k[t]
}{
\|\mathbf a(\widehat{\theta}_{k,c}^{\mathrm c})\|_2^2
}
\mathbf d_k^{*}(\rho)
\right|^2 .
\label{eq:ref_delay_search}
\end{equation}
The common AoA is then refined over
$G_{\theta}^{\mathrm{loc}}$ uniformly spaced points within
$\pm\Delta_{\theta}^{\mathrm{loc}}$ of
$\widehat{\theta}_{k,c}^{\mathrm c}$ by maximizing the
aggregated coherent power across snapshots as
\[
\begin{aligned}
\widehat{\theta}_{k,c}
={}&\arg\max_{\theta}\sum_{t=1}^{T}
\bigl|\mathbf a^{\mathsf H}(\theta)\mathbf Y_k[t] \operatorname{Diag}
\{\mathbf d_k^{*}(\widehat{\rho}_{k,c}[t])\}
\mathbf 1_{N_k}\bigr|^2,
\end{aligned}
\]
yielding
$\{(\widehat{\theta}_{k,c},
\widehat{\rho}_{k,c}[t])\}_{t=1}^{T}$.

For each AoA-refined candidate $c$, the reference-path response is
$\widehat c_{k,c}[t]
=
\mathbf a^{\mathsf H}(\widehat{\theta}_{k,c})
\mathbf Y_k[t]
\mathbf d_k^{*}(\widehat{\rho}_{k,c}[t])/
\|\mathbf a(\widehat{\theta}_{k,c})\|_2^2$.
The reliable candidate maximizing
$T^{-1}\sum_{t=1}^{T}|\widehat c_{k,c}[t]|^2$
is selected, yielding
$\widehat{\theta}_{k,\mathrm{ref}}$ and $\widehat{\rho}_k[t]$.
For the selected reference,
$\widehat c_k[t]
\approx
N_k\alpha_{k,\mathrm{ref}}e^{j\psi_k[t]}$,
where the phase of $\alpha_{k,\mathrm{ref}}$ is snapshot invariant and
$\psi_k[t]$ contains the snapshot-dependent phase introduced by the CFO and PO.
Accordingly,
$\widehat{\psi}_k[t]=\angle\widehat c_k[t]$.

The calibrated observation is then written as
\begin{equation}
\widetilde{\mathbf Y}_k[t]
=
e^{-j\widehat{\psi}_k[t]}
\mathbf Y_k[t]
\operatorname{Diag}
\left\{
\mathbf d_k^{*}
\bigl(\widehat{\rho}_k[t]\bigr)
\right\},
\label{eq:offset_compensation}
\end{equation}
where delay compensation removes TO and maps the $\ell$th path to the LoS-referenced relative delay $\tau_{k,\ell}^{\mathrm{rel}}=\tau_{k,\ell}-\tau_{k,\mathrm{ref}}$, while phase compensation removes the common packet phase introduced by the CFO and PO. For nearby UEs in the same group, similar UE-BS distances lead to closely aligned LoS delays, allowing the LoS path to serve as a shared reference. Consequently, after subtracting the corresponding LoS delays, the relative delays $\tau_{k,\ell}^{\mathrm{rel}}$ still retain cross-UE correlation. %As the compensation does not affect the AoA, the resulting relative delay--AoA supports remain exploitable in the subsequent shared-support recovery.

After offset mitigation, the calibrated observations
can be represented over an overcomplete delay-AoA dictionary. Let
$\{(\bar{\tau}_p,\bar{\theta}_p)\}_{p=1}^{P}$ denote the discretized
delay-AoA domain,\footnote{The delay interval $[0,\tau_{\max}]$ and the AoA interval
$[-\pi/2,\pi/2]$ are uniformly discretized into $G_\tau$ and
$G_\theta$ grid points, respectively. Their Cartesian product contains
$P=G_\tau G_\theta$ delay-AoA pairs, indexed by $p=1,\ldots,P$.}
where the $p$th atom for UE $k$ is defined as
$\boldsymbol{\phi}_{k,p}
\triangleq
\mathbf{d}_k(\bar{\tau}_p)\otimes\mathbf{a}(\bar{\theta}_p)$.
The resulting dictionary is
$\boldsymbol{\Phi}_k
=
[\boldsymbol{\phi}_{k,1},\ldots,\boldsymbol{\phi}_{k,P}]$.
By stacking the vectorized CSI observations over the $T$ snapshots, we obtain
\begin{equation}
\widetilde{\mathbf{Y}}_k^{\mathrm v}
=
\boldsymbol{\Phi}_k\mathbf{W}_k
+
\widetilde{\mathbf{E}}_k^{\mathrm v},
\label{eq:2d_sparse_model}
\end{equation}
where
$\widetilde{\mathbf{Y}}_k^{\mathrm v}
=
[\operatorname{vec}(\widetilde{\mathbf{Y}}_k[1]),\ldots,
\operatorname{vec}(\widetilde{\mathbf{Y}}_k[T])]$, $\widetilde{\mathbf{E}}_k^{\mathrm v}$ denotes the corresponding vectorized noise matrix, and
$\mathbf{W}_k\in\mathbb{C}^{P\times T}$ is row sparse. If path $\ell$ is represented by atom $p$, its coefficient at snapshot $t$
follows
$[\mathbf{W}_k]_{p,t}
=
\alpha_{k,\ell}e^{j2\pi\nu_{k,\ell}t_t}$,
whereas rows associated with inactive delay-AoA atoms are zero. Hence, the
row support of $\mathbf{W}_k$ identifies the active delay-AoA components,
while their snapshot-wise coefficient evolution contains the Doppler
information. Residual synchronization errors that remain in
$\widetilde{\mathbf{Y}}_k^{\mathrm v}$ are explicitly modeled in Section~III-C and
refined in the subsequent sensing estimators.

\subsection{Reduced Delay-AoA Candidate Construction}

For both multiuser sensing estimators developed in this paper, directly operating on the full delay-AoA dictionary would incur high computational complexity. We therefore construct a common reduced candidate set to provide a compact delay-AoA search space for the subsequent Group-SBL and JMUSIC processing.
For a delay-AoA pair $(\tau,\theta)$, define the normalized matching response
$\mathcal J_k(\mathbf Z;\theta,\tau)
\triangleq
|\mathbf a^{\mathsf H}(\theta)\mathbf Z\mathbf d_k^{*}(\tau)|^2/
(\|\mathbf a(\theta)\|_2^2\|\mathbf d_k(\tau)\|_2^2)$.

\emph{1) Successive coarse screening:}
The coarse screening uses
$G_{\tau}^{\mathrm c}$ uniformly spaced delay grid points
$\{\tau_g^{\mathrm c}\}_{g=1}^{G_{\tau}^{\mathrm c}}$ over
$[0,\tau_{\max}]$ and
$G_{\theta}^{\mathrm c}$ uniformly spaced AoA grid points
$\{\theta_q^{\mathrm c}\}_{q=1}^{G_{\theta}^{\mathrm c}}$ over
$[-\pi/2,\pi/2]$. For UE $k$, initialize the  working observation as
$\mathbf Z_k^{(0)}[t]=\widetilde{\mathbf Y}_k[t]$.
Let $i$ denote the successive screening iteration index.
At iteration $i$, the dominant peak of the current working observation
is selected as
\begin{equation}
(\widehat q_{k,i},\widehat g_{k,i})
=
\arg\max_{q,g}
\frac{1}{T}\sum_{t=1}^{T}
\mathcal J_k
\left(
\mathbf Z_k^{(i-1)}[t];
\theta_q^{\mathrm c},\tau_g^{\mathrm c}
\right).
\label{eq:residual_coarse_screening}
\end{equation}
The snapshot-wise least-squares (LS) coefficient is
\[
\begin{aligned}
\widehat{\alpha}_{k,i}[t]
={}&
\frac{
\mathbf a^{\mathsf H}(\theta_{\widehat q_{k,i}}^{\mathrm c})
\mathbf Z_k^{(i-1)}[t]
}{
\|\mathbf a(\theta_{\widehat q_{k,i}}^{\mathrm c})\|_2^2
}\times
\frac{
\mathbf d_k^{*}(\tau_{\widehat g_{k,i}}^{\mathrm c})
}{
\|\mathbf d_k(\tau_{\widehat g_{k,i}}^{\mathrm c})\|_2^2
},
\end{aligned}
\]
and the detected component is subtracted from the  observation as
\begin{equation}
\mathbf Z_k^{(i)}[t]
=
\mathbf Z_k^{(i-1)}[t]
-\widehat{\alpha}_{k,i}[t]
\mathbf a(\theta_{\widehat q_{k,i}}^{\mathrm c})
\mathbf d_k^{\mathsf T}(\tau_{\widehat g_{k,i}}^{\mathrm c}).
\end{equation}

Define $\mathcal PW_k^{(i)}\triangleq
\max_{q,g}T^{-1}\sum_{t=1}^{T}
\mathcal J_k(\mathbf Z_k^{(i)}[t];
\theta_q^{\mathrm c},\tau_g^{\mathrm c})$.
The screening terminates when
$\mathcal PW_k^{(i)}<\epsilon_{\mathrm c}\mathcal PW_k^{(0)}$
or $i=I_{\max}^{\mathrm c}$,
where $\epsilon_{\mathrm c}$ is a relative screening threshold
and $I_{\max}^{\mathrm c}$ is the maximum number of coarse-screening iterations.
The retained coarse-center set is then
$\mathcal C_k
=
\{(\tau_{\widehat g_{k,i}}^{\mathrm c},
\theta_{\widehat q_{k,i}}^{\mathrm c})\}_{i=1}^{I_k^{\mathrm c}}$, where $I_k^{\mathrm c}$ is the resulting number of screening iterations for UE $k$.

\emph{2) Local fine-grid selection:} The local fine-grid search is performed over the full delay-AoA grid
$G_{\tau}\times G_{\theta}$.
For the $i$th retained coarse center,
let $(q_{k,i}^{0},g_{k,i}^{0})$ denote its nearest index pair on the
full delay-AoA grid. A local fine-grid window is then constructed
around each coarse center. Define
$\overline{\mathcal J}_k(q,g)
\triangleq
\frac{1}{T}\sum_{t=1}^{T}
\mathcal J_k(\widetilde{\mathbf Y}_k[t];\theta_q,\tau_g)$.
The local window and the retained fine-grid candidates are 
\begin{equation}
\begin{aligned}
\mathcal B_k^{(i)}
&=
\left\{
(q,g):
|q-q_{k,i}^{0}|\leq r_{\theta},\
|g-g_{k,i}^{0}|\leq r_{\tau}
\right\},\\[-1mm]
\Omega_k^{(i)}
&=
\left\{
(\tau_g,\theta_q):
(q,g)\in
\operatorname{TopM}_{M_{\mathrm{loc}}}
\bigl(\overline{\mathcal J}_k(q,g),\mathcal B_k^{(i)}\bigr)
\right\},
\end{aligned}
\label{eq:local_fine_candidates}
\end{equation}
Here, $r_{\theta}=4$ and $r_{\tau}=2$ denote the local half-window
sizes in the AoA and delay dimensions, respectively, resulting in a
$(2r_{\theta}+1)\times(2r_{\tau}+1)$ local search window, and $\operatorname{TopM}_{M_{\mathrm{loc}}}$
selects the $M_{\mathrm{loc}}$ grid points with the largest
$\overline{\mathcal J}_k(q,g)$ values within
$\mathcal B_k^{(i)}$.
The UE-wise candidate set is thus
$\Omega_k=\bigcup_{i=1}^{I_k^{\mathrm c}}\Omega_k^{(i)}$.

Finally, the UE-wise sets are merged as
$\Omega=\bigcup_{k=1}^{K}\Omega_k$ with $P_\Omega=|\Omega|$,
and the corresponding columns of the full dictionary
$\boldsymbol{\Phi}_k$ form the reduced dictionary
$\boldsymbol{\Phi}_{k,\Omega}$.

\subsection{Residual Synchronization Error Model and Refinement}
\label{subsec:residual_sync}

The model in \eqref{eq:2d_sparse_model} represents the nominal
delay-AoA structure after the LoS-referenced offset mitigation.
In practice, the estimates
$\widehat{\rho}_k[t]$ and $\widehat{\psi}_k[t]$
in Section~III-A may be imperfect, leaving residual synchronization
errors in the calibrated observations.
Specifically, define the residual TO and residual common phase as
$\epsilon_k[t]\triangleq\rho_k[t]-\widehat{\rho}_k[t]$ and
$\xi_k[t]\triangleq\psi_k[t]-\widehat{\psi}_k[t]$, respectively.
The residual TO introduces a subcarrier-dependent phase distortion,
which is represented by
$\mathbf D_k(\epsilon)
=
\operatorname{Diag}
\left\{
e^{-j2\pi f_{k,1}\epsilon},
\ldots,
e^{-j2\pi f_{k,N_k}\epsilon}
\right\}
\otimes\mathbf I_M$.

Restricting \eqref{eq:2d_sparse_model} to the reduced candidate set $\Omega$, let
$\mathbf w_{k}[t]$ be the coefficient vector at snapshot $t$ and
$\widetilde{\mathbf y}_k[t]\triangleq
\operatorname{vec}(\widetilde{\mathbf Y}_k[t])$.
With $\epsilon_k[t]$ and $\xi_k[t]$, the calibrated observation in \eqref{eq:2d_sparse_model} becomes
\begin{equation}
\widetilde{\mathbf y}_k[t]
\approx
e^{j\xi_k[t]}
\mathbf D_k(\epsilon_k[t])
\boldsymbol{\Phi}_{k,\Omega}
\mathbf w_{k}[t]
+
\widetilde{\mathbf e}_k[t],
\label{eq:residual_sync_model}
\end{equation}
where $\widetilde{\mathbf e}_k[t]$ is the corresponding noise and
modeling error.

Based on \eqref{eq:residual_sync_model}, the joint refinement of the
residual TO and common phase can be formulated as
\begin{equation}
\min_{\epsilon_k[t],\,\xi_k[t],\,\mathbf w_{k}[t]}
\left\|
\widetilde{\mathbf y}_k[t]
\!-\!
e^{j\xi_k[t]}
\mathbf D_k(\epsilon_k[t])
\boldsymbol{\Phi}_{k,\Omega}
\mathbf w_{k}[t]
\right\|_2^2 .
\label{eq:general_residual_sync_problem}
\end{equation}

The two residual terms in \eqref{eq:general_residual_sync_problem} are handled differently. Since $\mathbf w_k[t]$ is complex and snapshot dependent, the factor $e^{j\xi_k[t]}$ can be absorbed into $\mathbf w_k[t]$ during delay-AoA estimation. We therefore defer the estimation of $\xi_k[t]$ to Section~VI-B, where it is recovered from the quasi-static reference path after the delay-AoA supports are obtained.

For the $\epsilon_k[t]$, define the centered subcarrier frequencies as
$\bar f_{k,n}\triangleq f_{k,n}-f_{k,\mathrm c}$, where
$f_{k,\mathrm c}\triangleq N_k^{-1}\sum_{n=1}^{N_k}f_{k,n}$.
Then, $\mathbf D_k(\epsilon)
=
e^{-j2\pi f_{k,\mathrm c}\epsilon}
\mathbf D_k^{\mathrm c}(\epsilon),$
where $\mathbf D_k^{\mathrm c}(\epsilon)
=
\operatorname{Diag}
\left\{
e^{-j2\pi\bar f_{k,1}\epsilon},
\ldots,
e^{-j2\pi\bar f_{k,N_k}\epsilon}
\right\}
\otimes\mathbf I_M .$
The scalar phase $e^{-j2\pi f_{k,\mathrm c}\epsilon}$ can also be absorbed into $\mathbf w_k[t]$, whereas $\mathbf D_k^{\mathrm c}(\epsilon)$ captures the relative phase variation across subcarriers and therefore contains the identifiable residual-TO information. Since $\epsilon_k[t]$ and $\mathbf w_k[t]$ are coupled in \eqref{eq:general_residual_sync_problem}, we adopt an \emph{alternating local update to iteratively refine them}. The following develops the common residual-TO update used by both
delay-AoA estimators.
%with their specific implementations given in Sections~IV-C and V-C.

 At an estimator iteration $i$ where residual-TO refinement is performed, let
$\widetilde{\mathbf Y}_{k,\mathrm{RTO}}^{(i-1)}[t]$
denote the residual-TO-compensated observation at snapshot $t$, with
$\widetilde{\mathbf y}_{k,\mathrm{RTO}}^{(i-1)}[t]
\triangleq
\operatorname{vec}(\widetilde{\mathbf Y}_{k,\mathrm{RTO}}^{(i-1)}[t])$.
Stacking the $T$ snapshots gives
$\widetilde{\mathbf Y}_{k,\mathrm{RTO}}^{\mathrm v(i-1)}
=
[\widetilde{\mathbf y}_{k,\mathrm{RTO}}^{(i-1)}[1],\ldots,
\widetilde{\mathbf y}_{k,\mathrm{RTO}}^{(i-1)}[T]]$.
The corresponding delay-AoA estimator first obtains
$\widehat{\mathbf w}_k^{(i)}[t]$ from
$\widetilde{\mathbf y}_{k,\mathrm{RTO}}^{(i-1)}[t]$ and forms
$\widehat{\mathbf s}_k^{(i)}[t]
=
\boldsymbol{\Phi}_{k,\Omega}
\widehat{\mathbf w}_k^{(i)}[t]$.
Conditioned on this reconstruction, the remaining residual-TO increment $\Delta\epsilon_k[t]$ is modeled as
\begin{equation}
\widetilde{\mathbf y}_{k,\mathrm{RTO}}^{(i-1)}[t]
\simeq
\mathbf D_k^{\mathrm c}(\Delta\epsilon_k[t])
\widehat{\mathbf s}_k^{(i)}[t]
+
\mathbf v_k^{(i)}[t],
\label{eq:generic_residual_to_model}
\end{equation}
where $\mathbf v_k^{(i)}[t]$ denotes the fitting error, including noise and reconstruction mismatch.
For a sufficiently small $\Delta\epsilon_k[t]$, $\mathbf D_k^{\mathrm c}(\Delta\epsilon_k[t])$
admits the first-order approximation $\mathbf D_k^{\mathrm c}(\Delta\epsilon_k[t])
\simeq
\mathbf I
+
\Delta\epsilon_k[t]\mathbf G_k^{\mathrm c},$
where
$\mathbf G_k^{\mathrm c}
=
\operatorname{Diag}
\left\{
-j2\pi\bar f_{k,1},
\ldots,
-j2\pi\bar f_{k,N_k}
\right\}
\otimes\mathbf I_M$.
Substituting the first-order approximation $\mathbf D_k^{\mathrm c}(\Delta\epsilon_k[t])$  into
\eqref{eq:generic_residual_to_model} gives
$\widetilde{\mathbf y}_{k,\mathrm{RTO}}^{(i-1)}[t]
\simeq
\widehat{\mathbf s}_k^{(i)}[t]
+
\mathbf G_k^{\mathrm c}
\widehat{\mathbf s}_k^{(i)}[t]
\Delta\epsilon_k[t]
+
\mathbf v_k^{(i)}[t]$.

Defining $\mathbf r_k^{(i)}[t]
\triangleq
\widetilde{\mathbf y}_{k,\mathrm{RTO}}^{(i-1)}[t]
-
\widehat{\mathbf s}_k^{(i)}[t],
\mathbf b_k^{(i)}[t]
\triangleq
\mathbf G_k^{\mathrm c}
\widehat{\mathbf s}_k^{(i)}[t],$
the residual-TO refinement reduces to
$\mathbf r_k^{(i)}[t]
\simeq
\mathbf b_k^{(i)}[t]\Delta\epsilon_k[t]
+
\mathbf v_k^{(i)}[t]$.
The local fitting error is modeled as circular complex Gaussian,
$\mathbf v_k^{(i)}[t]
\sim
\mathcal{CN}
\left(
\mathbf 0,
(\kappa_k^{(i)})^{-1}\mathbf I
\right)$,
with $\kappa_k^{(i)}$ denoting the corresponding error precision.
We further impose a zero-mean Gaussian prior of the form
$p(\Delta\epsilon)\propto\exp(-\lambda_\epsilon\Delta\epsilon^2)$,
where $\lambda_\epsilon>0$ controls the regularization strength.
The corresponding maximum a posteriori (MAP) estimate is therefore\cite{Stuart2010}
\begin{equation}
\widehat{\Delta\epsilon}_k[t]
\!=\!
\arg\min_{\Delta\epsilon\in\mathbb R}
\left\{\!
\kappa_k^{(i)}
\left\|
\mathbf r_k^{(i)}[t]
\!-\!
\mathbf b_k^{(i)}[t]\Delta\epsilon
\right\|_2^2
\!\!+\!\!
\lambda_\epsilon\Delta\epsilon^2
\!\right\}.
\label{eq:residual_to_subproblem}
\end{equation}
Since the objective in \eqref{eq:residual_to_subproblem} is a convex
quadratic function of the real-valued $\Delta\epsilon$, setting its derivative to zero gives
\begin{equation}
\widehat{\Delta\epsilon}_k[t]
=
\operatorname{clip}_{\Delta\epsilon_{\max}}
\left(
\frac{
\kappa_k^{(i)}
\operatorname{Re}
\left\{
\left(\mathbf b_k^{(i)}[t]\right)^{\mathsf H}
\mathbf r_k^{(i)}[t]
\right\}
}{
\kappa_k^{(i)}
\left\|\mathbf b_k^{(i)}[t]\right\|_2^2
+
\lambda_\epsilon
}
\right),
\label{eq:residual_to_update}
\end{equation}
where
$\operatorname{clip}_{a}(x)\triangleq\max\{-a,\min\{a,x\}\}$
and $\Delta\epsilon_{\max}>0$ denotes the maximum allowable residual-TO increment in each update.

After estimating the residual-TO increments over all $T$ snapshots, a snapshot-invariant timing ambiguity may remain. To remove this ambiguity, we normalize the estimated increments to have zero mean,
\begin{equation}
\widehat{\Delta\epsilon}_k[t]
\leftarrow
\widehat{\Delta\epsilon}_k[t]
-
\frac{1}{T}
\sum_{t'=1}^{T}
\widehat{\Delta\epsilon}_k[t'],
t=1,\ldots,T.
\label{eq:residual_to_gauge}
\end{equation}
The normalized increments are then accumulated with the previous
residual-TO estimate as
\begin{equation}
\widehat{\epsilon}_{k,\mathrm{RTO}}^{(i)}[t]
=
\widehat{\epsilon}_{k,\mathrm{RTO}}^{(i-1)}[t]
+
\widehat{\Delta\epsilon}_k[t].
\label{eq:residual_to_accumulation}
\end{equation}
The accumulated residual-TO estimate
$\widehat{\epsilon}_{k,\mathrm{RTO}}^{(i)}[t]$
is then used to update the residual-TO-compensated observation for the
subsequent delay-AoA estimation. The estimator-specific constructions of
$\widehat{\mathbf s}_k^{(i)}[t]$, $\kappa_k^{(i)}$, together with the
update of $\widetilde{\mathbf Y}_{k,\mathrm{RTO}}^{\mathrm v(i)}$, are detailed in Sections~IV-C and V-B. %The residual common phase $\xi_k[t]$ is treated separately in Section~VI-B.

\section{Group Shared-Private SBL}
As the first multiuser sensing estimator, we develop a latent-group
shared-private SBL scheme (Group-SBL) that explicitly exploits cross-UE
delay-AoA correlation through group-level shared sparsity. Its variational inference is periodically alternated with residual-TO refinement to jointly improve the delay-AoA estimation and synchronization accuracy.
\subsection{Shared-Private Bayesian Model}
Building on the per-UE sparse model in \eqref{eq:2d_sparse_model},
the coefficient matrix $\mathbf W_k$ is decomposed as
$\mathbf{W}_k=\mathbf{W}_k^{\mathrm{sh}}+\mathbf{W}_k^{\mathrm{pr}}$ \cite{ref_dai_grouping},
where $\mathbf{W}_k^{\mathrm{sh}}$ captures delay-AoA atoms shared within a
propagation group, while $\mathbf{W}_k^{\mathrm{pr}}$ absorbs non-common
paths. 
This formulation exploits shared supports across UEs while accommodating UE-specific variations to improve more reliable support recovery.
%This formulation consolidates evidence for shared supports across UEs while accommodating UE-independent variations, resulting in more reliable support recovery. %Once the delay-AoA supports are determined, their temporal coefficient sequences can form a reduced one-dimensional model for Doppler estimation.

Based on this shared-private decomposition and the reduced candidate set
$\Omega$, define
$\overline{\boldsymbol{\Phi}}_{k,\Omega}
\triangleq
[\boldsymbol{\Phi}_{k,\Omega},
 \boldsymbol{\Phi}_{k,\Omega}]$,
and the corresponding multiple-measurement-vector (MMV) model is
\begin{equation}
\widetilde{\mathbf Y}_{k}^{\mathrm v}
=
\overline{\boldsymbol{\Phi}}_{k,\Omega}
\overline{\mathbf W}_k
+
\widetilde{\mathbf E}_{k}^{\mathrm v},
\qquad
\overline{\mathbf W}_k
=
\begin{bmatrix}
\mathbf W_k^{\mathrm{sh}}\\
\mathbf W_k^{\mathrm{pr}}
\end{bmatrix},
\label{eq:reduced_shared_private_model}
\end{equation}
where
$\mathbf W_k^{\mathrm{sh}},
\mathbf W_k^{\mathrm{pr}}
\in\mathbb C^{P_{\Omega}\times T}$
denote the group-shared and UE-specific coefficient matrices,
respectively.

To characterize the latent group structure, we consider $C$ candidate
UE groups. Let
$\mathbf z_k=[z_{k,1},\ldots,z_{k,C}]^{\mathsf T}$ denote the group
indicator of UE $k$, where $z_{k,c}\in\{0,1\}$ with $z_{k,c}=1$ indicating membership in group $c$, and
$\sum_c z_{k,c}=1$. Given the group proportions
$\boldsymbol{\pi}=[\pi_1,\ldots,\pi_C]^{\mathsf T}$, we assume
$p(\boldsymbol{\pi})=\operatorname{Dir}(\boldsymbol{\pi};
\boldsymbol{\alpha}_0)$ and
$p(\mathbf z_k\mid\boldsymbol{\pi})
=\prod_{c=1}^{C}\pi_c^{z_{k,c}}$.

Let $\mathbf w_{k,p}^{\mathrm{sh}}$ and
$\mathbf w_{k,p}^{\mathrm{pr}}$ denote the $p$th rows of
$\mathbf W_k^{\mathrm{sh}}$ and $\mathbf W_k^{\mathrm{pr}}$,
respectively. Their sparsity-inducing priors are
\begin{equation}
\begin{aligned}
p(\mathbf w_{k,p}^{\mathrm{sh}}
 \mid\mathbf z_k,\{\gamma_{p,c}\})
&=
\prod_{c=1}^{C}
\mathcal{CN}
\left(
\mathbf w_{k,p}^{\mathrm{sh}};
\mathbf 0,\gamma_{p,c}^{-1}\mathbf I_T
\right)^{z_{k,c}},\\[-1mm]
p(\mathbf w_{k,p}^{\mathrm{pr}}\mid\eta_{k,p})
&=
\mathcal{CN}
\left(
\mathbf w_{k,p}^{\mathrm{pr}};
\mathbf 0,
\rho_{\mathrm{pr}}\eta_{k,p}^{-1}\mathbf I_T
\right).
\end{aligned}
\label{eq:shared_private_priors}
\end{equation}
where $\gamma_{p,c}$ and $\eta_{k,p}$ are sparsity precision
parameters for the shared and private components, respectively. Specifically,
$\gamma_{p,c}$ is shared by the UEs assigned to group $c$ and controls the
group-level activation of atom $p$, whereas $\eta_{k,p}$ governs its
UE-specific activation. The factor $\rho_{\mathrm{pr}}=0.001$ strongly promotes private-component sparsity, thereby encouraging jointly supported atoms to be explained by the shared component.
%imposes a stronger sparsity preference on the private component, thereby encouraging jointly supported atoms to be explained by the shared component.

Let $\widetilde{\mathbf e}_k[t]$ denote the $t$th column of
$\widetilde{\mathbf E}_k^{\mathrm v}$. We model
$\widetilde{\mathbf e}_k[t]\sim
\mathcal{CN}(\mathbf 0,\zeta_k^{-1}\mathbf I_{D_k})$,
where $\zeta_k=\sigma_{e,k}^{-2}$ denotes the noise precision. Independent Gamma hyperpriors are assigned as
$p(\gamma_{p,c})
=\operatorname{Ga}(\gamma_{p,c};a_{\gamma0},b_{\gamma0})$,
$p(\eta_{k,p})
=\operatorname{Ga}(\eta_{k,p};a_{\eta0},b_{\eta0})$, and
$p(\zeta_k)
=\operatorname{Ga}(\zeta_k;a_{\zeta0},b_{\zeta0})$, where
$\operatorname{Ga}(\cdot)$ follows the shape-rate convention.
With $D_k=MN_k$, the likelihood is
$p(\widetilde{\mathbf Y}_{k}^{\mathrm v}
\mid\overline{\mathbf W}_k,\zeta_k)
=
\mathcal{CN}(
\widetilde{\mathbf Y}_{k}^{\mathrm v};
\overline{\boldsymbol{\Phi}}_{k,\Omega}\overline{\mathbf W}_k,
\zeta_k^{-1}\mathbf I_{D_k})$. 

\subsection{Variational Inference}

Based on the hierarchical model above, the unknown quantities include
the augmented coefficient matrices
$\{\overline{\mathbf W}_k\}_{k=1}^{K}$, UE-group indicators
$\{\mathbf z_k\}_{k=1}^{K}$, group-proportion vector
$\boldsymbol{\pi}$, shared sparsity precisions
$\{\gamma_{p,c}\}_{p=1,c=1}^{P_{\Omega},C}$, private sparsity
precisions $\{\eta_{k,p}\}_{k=1,p=1}^{K,P_{\Omega}}$, and noise
precisions $\{\zeta_k\}_{k=1}^{K}$.
The shared precisions $\{\gamma_{p,c}\}$ aggregate posterior evidence
across UEs through the responsibilities $\{r_{k,c}\}$, 
reinforcing delay-AoA atoms commonly supported within the same
propagation group, whereas $\{\eta_{k,p}\}$ preserve UE-specific
components.

Given the calibrated observations
$\{\widetilde{\mathbf Y}_k^{\mathrm v}\}_{k=1}^{K}$, the joint posterior
of these quantities is analytically intractable due to the latent
UE-group assignments and hierarchical sparsity model. We therefore
adopt the mean-field variational Bayesian procedure in
\cite{ref_dai_grouping}, with posterior factors
$q(\overline{\mathbf W}_k)$, $q(\mathbf z_k)$,
$q(\boldsymbol{\pi})$, $q(\gamma_{p,c})$,
$q(\eta_{k,p})$, and $q(\zeta_k)$.
These factors are updated sequentially to approximate the joint
posterior as follows:

\emph{a) Update of $q(\overline{\mathbf W}_k)$:}
Let
$r_{k,c}\triangleq q(z_{k,c}=1)$,
$\lambda_{k,p}^{\mathrm{sh}}
=\sum_{c=1}^{C}r_{k,c}\mathbb E[\gamma_{p,c}]$, and
$\lambda_{k,p}^{\mathrm{pr}}
=\mathbb E[\eta_{k,p}]/\rho_{\mathrm{pr}}$.
Define
$\boldsymbol{\Lambda}_k
=\operatorname{Diag}
(\{\lambda_{k,p}^{\mathrm{sh}}\}_{p=1}^{P_{\Omega}},
 \{\lambda_{k,p}^{\mathrm{pr}}\}_{p=1}^{P_{\Omega}})$.
The coefficient posterior is complex Gaussian with covariance
$\boldsymbol{\Sigma}_k$ and mean $\mathbf U_k$:
\begin{equation}
\boldsymbol{\Sigma}_k\!=\!
\bigl(\!
\mathbb E[\zeta_k]
\overline{\boldsymbol{\Phi}}_{k,\Omega}^{\mathsf H}
\overline{\boldsymbol{\Phi}}_{k,\Omega}
\!+\!\boldsymbol{\Lambda}_k
\bigr)^{-1},
\mathbf U_k\!=\!
\mathbb E[\zeta_k]\boldsymbol{\Sigma}_k
\overline{\boldsymbol{\Phi}}_{k,\Omega}^{\mathsf H}
\widetilde{\mathbf Y}_{k}^{\mathrm v}.
\label{eq:coefficient_posterior_update}
\end{equation}
Partition
$\mathbf U_k
=[(\mathbf U_k^{\mathrm{sh}})^{\mathsf T},
(\mathbf U_k^{\mathrm{pr}})^{\mathsf T}]^{\mathsf T}$,
with posterior row energies
$S_{k,p}^{\mathrm{sh}}
=\|\mathbf U_k^{\mathrm{sh}}(p,:)\|_2^2
+T[\boldsymbol{\Sigma}_k]_{p,p}$
and
$S_{k,p}^{\mathrm{pr}}
=\|\mathbf U_k^{\mathrm{pr}}(p,:)\|_2^2
+T[\boldsymbol{\Sigma}_k]_{P_{\Omega}+p,P_{\Omega}+p}$.

\emph{b) Updates of $q(\gamma_{p,c})$ and $q(\eta_{k,p})$:}
Both precision posteriors remain Gamma distributed as
$q(\gamma_{p,c})
=\operatorname{Ga}(\gamma_{p,c};
a_{p,c}^{\gamma},b_{p,c}^{\gamma})$
and
$q(\eta_{k,p})
=\operatorname{Ga}(\eta_{k,p};
a_{k,p}^{\eta},b_{k,p}^{\eta})$,
where their parameters are updated as
$a_{p,c}^{\gamma}=a_{\gamma0}+T\sum_{k=1}^{K}r_{k,c}$,
$b_{p,c}^{\gamma}=b_{\gamma0}+\sum_{k=1}^{K}r_{k,c}S_{k,p}^{\mathrm{sh}}$,
$a_{k,p}^{\eta}=a_{\eta0}+T$, and
$b_{k,p}^{\eta}=b_{\eta0}+\rho_{\mathrm{pr}}^{-1}S_{k,p}^{\mathrm{pr}}$.
Hence,
$\mathbb E[\gamma_{p,c}]
=a_{p,c}^{\gamma}/b_{p,c}^{\gamma}$ and
$\mathbb E[\eta_{k,p}]
=a_{k,p}^{\eta}/b_{k,p}^{\eta}$.
The shared-precision update combines the posterior evidence of
multiple UEs through $r_{k,c}$, thereby reinforcing commonly
supported delay-AoA atoms.

\emph{c) Updates of $q(\boldsymbol{\pi})$ and $q(\mathbf z_k)$:}
The group-proportion posterior is
$q(\boldsymbol{\pi})
=\operatorname{Dir}(\boldsymbol{\pi};\boldsymbol{\alpha})$, with
$\alpha_c=\alpha_{0,c}+\sum_{k=1}^{K}r_{k,c}$.
The responsibility is updated as
\begin{equation}
r_{k,c}=
{\exp(\xi_{k,c})}/
{\sum_{c'=1}^{C}\exp(\xi_{k,c'})},
\label{eq:group_responsibility_update}
\end{equation}
where
$\xi_{k,c}
=
\mathbb E[\log\pi_c]
+
\sum_{p=1}^{P_{\Omega}}
\left[
T\mathbb E[\log\gamma_{p,c}]
-
\mathbb E[\gamma_{p,c}]S_{k,p}^{\mathrm{sh}}
\right]$,
$\mathbb E[\log\pi_c]
=\psi(\alpha_c)-\psi(\sum_{c'}\alpha_{c'})$,
and
$\mathbb E[\log\gamma_{p,c}]
=\psi(a_{p,c}^{\gamma})-\log b_{p,c}^{\gamma}$,
with $\psi(\cdot)$ denoting the digamma function.
Thus, the shared-support estimates and UE grouping are alternately
reinforced.

\emph{d) Update of $q(\zeta_k)$:}
Let
$\mathcal E_k
\triangleq
\|\widetilde{\mathbf Y}_{k}^{\mathrm v}
-\overline{\boldsymbol{\Phi}}_{k,\Omega}\mathbf U_k\|_{\mathrm F}^{2}
+T\operatorname{tr}
(\boldsymbol{\Sigma}_k
\overline{\boldsymbol{\Phi}}_{k,\Omega}^{\mathsf H}
\overline{\boldsymbol{\Phi}}_{k,\Omega})$.
The noise-precision posterior is
$q(\zeta_k)
=
\operatorname{Ga}(\zeta_k;a_k^{\zeta},b_k^{\zeta})$,
where
$a_k^{\zeta}=a_{\zeta0}+TD_k$,
$b_k^{\zeta}=b_{\zeta0}+\mathcal E_k$, and
$\mathbb E[\zeta_k]=a_k^{\zeta}/b_k^{\zeta}$.

The resulting posterior mean $\mathbf U_k$ and noise-precision
posterior $q(\zeta_k)$ are subsequently used to instantiate the
residual-TO refinement in Section~IV-C.
\subsection{Posterior-Aided Residual-TO Refinement}

The residual-TO update in Section~\ref{subsec:residual_sync} requires
$\widehat{\mathbf s}_k^{(i)}[t]$ and $\kappa_k^{(i)}$.
For Group-SBL, both are obtained from the current variational posterior. The residual-TO-compensated MMV observation is initialized as
$\widetilde{\mathbf Y}_{k,\mathrm{RTO}}^{\mathrm v(0)}
=
\widetilde{\mathbf Y}_k^{\mathrm v}$,
with
$\widehat{\epsilon}_{k,\mathrm{RTO}}^{(0)}[t]=0$.
To allow the SBL posterior to become sufficiently stabilized before residual-TO refinement, the first TO update is performed at SBL iteration $I_{\mathrm{TO}}$ and subsequently every $P_{\mathrm{TO}}$ iterations, thereby periodically alternating residual-TO refinement with the SBL updates for at most $N_{\mathrm{TO}}^{\max}$ update rounds. Accordingly,
$\mathcal I_{\mathrm{TO}}
\triangleq
\{I_{\mathrm{TO}}+mP_{\mathrm{TO}}:
m=0,\ldots,N_{\mathrm{TO}}^{\max}-1\}$.

For $i\in\mathcal I_{\mathrm{TO}}$, the current posterior means give the
reconstructed propagation signal
\begin{equation}
\widehat{\mathbf Y}_k^{\mathrm v(i)}
=
\boldsymbol{\Phi}_{k,\Omega}
(
\mathbf U_k^{\mathrm{sh}(i)}
+
\mathbf U_k^{\mathrm{pr}(i)}
),
\label{eq:residual_to_reconstruction}
\end{equation}
where its $t$th column
$\widehat{\mathbf y}_k^{(i)}[t]$
provides the reconstruction required in Section~\ref{subsec:residual_sync},
i.e.,
$\widehat{\mathbf s}_k^{(i)}[t]
=
\widehat{\mathbf y}_k^{(i)}[t]$.
The corresponding local fitting-error precision is obtained from the
noise-precision posterior in Section~IV-B as
$\kappa_k^{(i)}=\mathbb E[\zeta_k]$.

Using these quantities in the general update of
Section~\ref{subsec:residual_sync}, we have
$\mathbf r_k^{(i)}[t]
=
\widetilde{\mathbf y}_{k,\mathrm{RTO}}^{(i-1)}[t]
-
\widehat{\mathbf y}_k^{(i)}[t]$
and
$\mathbf b_k^{(i)}[t]
=
\mathbf G_k^{\mathrm c}
\widehat{\mathbf y}_k^{(i)}[t]$.
The residual-TO increment
$\widehat{\Delta\epsilon}_k[t]$
is therefore obtained directly from
\eqref{eq:residual_to_update}, followed by the zero-mean normalization
in \eqref{eq:residual_to_gauge} and the accumulation in
\eqref{eq:residual_to_accumulation}.

The accumulated residual-TO estimate is then applied to the
frontend-calibrated observation as
\begin{equation}
\widetilde{\mathbf Y}_{k,\mathrm{RTO}}^{(i)}[t]
=
\widetilde{\mathbf Y}_k[t]
\operatorname{Diag}
\{
e^{j2\pi\bar f_{k,n}
\widehat{\epsilon}_{k,\mathrm{RTO}}^{(i)}[t]}
\}_{n=1}^{N_k}.
\label{eq:centered_rto_compensation}
\end{equation}
The updated observations are then vectorized and stacked to form
$\widetilde{\mathbf Y}_{k,\mathrm{RTO}}^{\mathrm v(i)}$, which is used
in the subsequent variational updates.
The complete iterative procedure is summarized in
Algorithm~\ref{alg:group_sbl_2d}. The SBL convergence is measured by
$\Delta_U^{(i)}
=
\sum_{k=1}^{K}
\|\mathbf U_k^{(i)}-\mathbf U_k^{(i-1)}\|_{\mathrm F}^{2}
/\sum_{k=1}^{K}
\|\mathbf U_k^{(i-1)}\|_{\mathrm F}^{2}$,
and the iterations terminate when
$\Delta_U^{(i)}<\varepsilon_{\mathrm{VI}}$
or when $I_{\max}$ is reached. 
Moreover, $\widehat L_k$ dominant delay-AoA components are retained from
$\|\mathbf U_k^{\mathrm{sh}}(p,:)\|_2^2+
\|\mathbf U_k^{\mathrm{pr}}(p,:)\|_2^2$,
yielding $\{\{(\widehat{\tau}_{k,\ell},
\widehat{\theta}_{k,\ell})\}_{\ell=1}^{\widehat L_k}\}_{k=1}^{K}$.

\begin{algorithm}[t]
\caption{Shared-private 2D SBL}
\label{alg:group_sbl_2d}
\begin{algorithmic}[1]

\REQUIRE $\{\widetilde{\mathbf Y}_{k}^{\mathrm v},\boldsymbol{\Phi}_{k,\Omega}\}_{k=1}^{K}$,
$C$, $I_{\max}$, $\varepsilon_{\mathrm{VI}}$,
$I_{\mathrm{TO}}$, $P_{\mathrm{TO}}$,
$N_{\mathrm{TO}}^{\max}$.

\STATE Initialize $\{r_{k,c}\}$, $\{\mathbb E[\gamma_{p,c}]\}$,
$\{\mathbb E[\eta_{k,p}]\}$, $\{\mathbb E[\zeta_k]\}$,
$\widehat{\epsilon}_{k,\mathrm{RTO}}^{(0)}[t]=0$,
$\widetilde{\mathbf Y}_{k,\mathrm{RTO}}^{\mathrm v(0)}
=\widetilde{\mathbf Y}_{k}^{\mathrm v}$.

\FOR{$i=1,\ldots,I_{\max}$}

   \STATE Update
$q(\overline{\mathbf W}_k)$,
$q(\gamma_{p,c})$,
$q(\eta_{k,p})$,
$q(\boldsymbol{\pi})$,
$q(\mathbf z_k)$,
$q(\zeta_k)$ from
$\widetilde{\mathbf Y}_{k,\mathrm{RTO}}^{\mathrm v(i-1)}$
via \emph{a-d} in Section IV-B.
\IF{$i\in\mathcal I_{\mathrm{TO}}$}
    \FOR{$k=1,\ldots,K$}

        \STATE Obtain
        $\widehat{\mathbf Y}_k^{\mathrm v(i)}$
        from \eqref{eq:residual_to_reconstruction}.

        \FOR{$t=1,\ldots,T$}
            \STATE Form
            $\mathbf r_k^{(i)}[t], \mathbf b_k^{(i)}[t]$ and
            obtain $\widehat{\Delta\epsilon}_k[t]$ via
            \eqref{eq:residual_to_update}.
        \ENDFOR

       \STATE Normalize and accumulate
$\{\widehat{\Delta\epsilon}_k[t]\}_{t=1}^{T}$
via \eqref{eq:residual_to_gauge} and
\eqref{eq:residual_to_accumulation}; update the observations via
\eqref{eq:centered_rto_compensation}, then vectorize and stack them
to form
$\widetilde{\mathbf Y}_{k,\mathrm{RTO}}^{\mathrm v(i)}$.

    \ENDFOR
    \ELSE
    \STATE
    $\widehat{\epsilon}_{k,\mathrm{RTO}}^{(i)}[t]
    =
    \widehat{\epsilon}_{k,\mathrm{RTO}}^{(i-1)}[t]$ and
    $\widetilde{\mathbf Y}_{k,\mathrm{RTO}}^{\mathrm v(i)}
    =
    \widetilde{\mathbf Y}_{k,\mathrm{RTO}}^{\mathrm v(i-1)}$.
\ENDIF

\STATE \textbf{if} $\Delta_U^{(i)}<\varepsilon_{\mathrm{VI}}$ \textbf{then break}
\ENDFOR

\textbf{Output:} The final residual-TO-refined observations
$\{\widetilde{\mathbf Y}_{k}^{\mathrm v\star}\}_{k=1}^{K}$
and
$\{\{(\widehat{\tau}_{k,\ell},\widehat{\theta}_{k,\ell})\}_{\ell=1}^{\widehat L_k}\}_{k=1}^{K}$.

\end{algorithmic}
\end{algorithm}

\section{Joint MUSIC}
\label{sec:joint_music}
The preceding Group-SBL method exploits cross-UE delay-AoA
correlation through an explicit shared-private probabilistic model,
directly recovering UE-specific supports at the
cost of iterative Bayesian inference.
As an alternative deterministic approach, we develop a
JMUSIC sensing estimator that implicitly exploits cross-UE
correlation through joint subspace processing.
The calibrated UE observations are combined to form a joint
signal subspace, where the reduced delay-AoA candidates are evaluated to
construct a global pool encompassing the delay-AoA components of all UEs.
This pool directly supports scene-level sensing
without UE-specific association, but its reliability depends
on accurate subspace estimation and appropriate model-order selection.
We further develop a concentrated DML-based UE association
with residual-TO refinement to recover UE-specific delay-AoA supports.

\subsection{Multiuser Subspace and Global Delay-AoA Pool}
\label{subsec:joint_music_global_pool}

\subsubsection{Stacked Multiuser Observation and Candidate Blocks}

Recall from Section~III-B that
$\Omega=\{(\bar{\tau}_p,\bar{\theta}_p)\}_{p=1}^{P_\Omega}$
denotes the common reduced candidate set. Restricting
\eqref{eq:2d_sparse_model} to $\Omega$ gives
$\widetilde{\mathbf Y}_k^{\mathrm v}
=\boldsymbol{\Phi}_{k,\Omega}\mathbf W_k
+\widetilde{\mathbf E}_k^{\mathrm v}$,
where
$\boldsymbol{\Phi}_{k,\Omega}
=[\boldsymbol{\phi}_{k,1},\ldots,\boldsymbol{\phi}_{k,P_\Omega}]$,
$\boldsymbol{\phi}_{k,p}
=\mathbf d_k(\bar{\tau}_p)\otimes\mathbf a(\bar{\theta}_p)$,
and $D_k=MN_k$.

To aggregate multiuser observations, we first remove
the estimated LoS reference, located at zero relative
delay after calibration, and then stack the remaining observations. For UE $k$, define its estimated
reference atom as
$\widehat{\boldsymbol{\phi}}_{k,\mathrm{ref}}
=
\mathbf d_k(0)\otimes
\mathbf a(\widehat{\theta}_{k,\mathrm{ref}})
\in\mathbb C^{D_k}$.
Its snapshot-dependent complex coefficients are estimated by regularized
least squares as
$\widehat{\mathbf c}_{k,\mathrm{ref}}
=
\left(
\widehat{\boldsymbol{\phi}}_{k,\mathrm{ref}}^{\mathsf H}
\widehat{\boldsymbol{\phi}}_{k,\mathrm{ref}}
+\lambda_{\mathrm{LS}}
\right)^{-1}
\widehat{\boldsymbol{\phi}}_{k,\mathrm{ref}}^{\mathsf H}
\widetilde{\mathbf Y}_{k}^{\mathrm v}$,
where
$\widehat{\mathbf c}_{k,\mathrm{ref}}\in\mathbb C^{1\times T}$ and
$\lambda_{\mathrm{LS}}>0$ is a small regularization coefficient.
Hence, the reference-suppressed observation is
$\widetilde{\mathbf Y}_{k,\mathrm N}^{\mathrm v}
=\widetilde{\mathbf Y}_{k}^{\mathrm v}
-\widehat{\boldsymbol{\phi}}_{k,\mathrm{ref}}
\widehat{\mathbf c}_{k,\mathrm{ref}}$.

Stacking the reference-suppressed observations gives
\begin{equation}
\begin{aligned}
\widetilde{\mathbf Y}_{\mathrm J}
&\!=\!
[(\widetilde{\mathbf Y}_{1,\mathrm N}^{\mathrm v})^{\mathsf T},
..,
(\widetilde{\mathbf Y}_{K,\mathrm N}^{\mathrm v})^{\mathsf T}]^{\mathsf T}
\\
&\!=\!
[(\boldsymbol{\Phi}_{1,\Omega}\mathbf W_1)^{\mathsf T},
..,
(\boldsymbol{\Phi}_{K,\Omega}\mathbf W_K)^{\mathsf T}]^{\mathsf T}
\!+\!\widetilde{\mathbf E}_{\mathrm J}
\!\!\in\mathbb C^{D_{\mathrm J}\times T}.
\end{aligned}
\label{eq:joint_music_stacked_observation}
\end{equation}
where
$D_{\mathrm J}=\sum_{k=1}^{K}D_k$ and
$\widetilde{\mathbf E}_{\mathrm J}
\triangleq
[(\widetilde{\mathbf E}_{1}^{\mathrm v})^{\mathsf T},
\ldots,
(\widetilde{\mathbf E}_{K}^{\mathrm v})^{\mathsf T}]^{\mathsf T}$, $\mathbf W_k$ here represents the coefficients of the remaining
components after suppressing the estimated reference contribution.

Although all UE blocks are represented over the common reduced candidate
set $\Omega$, different UEs retain independent coefficients
$\mathbf W_k$ for the same candidate. We therefore represent each
delay-AoA candidate by a multiuser candidate block. Let
$\mathbf q_{k,p}
=
\boldsymbol{\phi}_{k,p}/\|\boldsymbol{\phi}_{k,p}\|_2$
denote the normalized response of UE $k$ to candidate $p$. This gives the multiuser candidate matrix
\begin{equation}
\mathbf B_p
=
\operatorname{blkdiag}
\left(
\mathbf q_{1,p},
\ldots,
\mathbf q_{K,p}
\right)
\in\mathbb C^{D_{\mathrm J}\times K},
\label{eq:joint_music_block_steering}
\end{equation}
which satisfies
$\mathbf B_p^{\mathsf H}\mathbf B_p=\mathbf I_K$. Accordingly, \eqref{eq:joint_music_stacked_observation} becomes
\begin{equation}
\widetilde{\mathbf Y}_{\mathrm J}
=
\sum_{p=1}^{P_\Omega}
\mathbf B_p\mathbf X_p
+
\widetilde{\mathbf E}_{\mathrm J},
\label{eq:joint_music_block_model}
\end{equation}
where
$\mathbf X_p
=
[
\|\boldsymbol{\phi}_{1,p}\|_2[\mathbf W_1]_{p,:}^{\mathsf T},
\ldots,
\|\boldsymbol{\phi}_{K,p}\|_2[\mathbf W_K]_{p,:}^{\mathsf T}
]^{\mathsf T}
\in\mathbb C^{K\times T}$.
Thus, $\operatorname{span}(\mathbf B_p)$ represents the multiuser subspace
associated with candidate $p$, while the rows of $\mathbf X_p$ retain the
independent contributions of the $K$ UEs.

\subsubsection{Signal-Subspace Construction}

Using the stacked multiuser observation, we next determine the
signal-subspace dimension.
A compact singular value decomposition (SVD) is first performed as
$\widetilde{\mathbf Y}_{\mathrm J}
=\mathbf U_{\mathrm J}\boldsymbol{\Sigma}_{\mathrm J}
\mathbf V_{\mathrm J}^{\mathsf H}
\in\mathbb C^{D_{\mathrm J}\times T}$.
Since $D_{\mathrm J}$ may be much larger than the number of CSI snapshots
$T$, the signal-subspace dimension is determined using the
high-dimensional source-enumeration method in
\cite{NadakuditiEdelman2008}. Let $s_i$ denote the singular values of
$\widetilde{\mathbf Y}_{\mathrm J}$ and
$\lambda_i=s_i^2/T$ the corresponding nonzero sample-covariance
eigenvalues, with the remaining sample-covariance eigenvalues equal to zero.

For a candidate dimension $d$, let $\beta_{\mathrm{NE}}=2$
for complex-valued observations and define
\[
\begin{aligned}
t_{D_{\mathrm J},d}
&=(D_{\mathrm J}-d)
\frac{\sum_{i=d+1}^{D_{\mathrm J}}\lambda_i^2}
{\left(\sum_{i=d+1}^{D_{\mathrm J}}\lambda_i\right)^2},\\
q_d
&=D_{\mathrm J}
\left[t_{D_{\mathrm J},d}
-\left(1+\frac{D_{\mathrm J}}{T}\right)\right]
-\left(\frac{2}{\beta_{\mathrm{NE}}}-1\right)
\frac{D_{\mathrm J}}{T},\\
d_{\mathrm s}
&=\operatorname*{arg\,min}_{1\leq d\leq d_{\max}}
\left\{
\frac{\beta_{\mathrm{NE}}}{4}
\left(\frac{T}{D_{\mathrm J}}\right)^2q_d^2
+2(d+1)
\right\},
\end{aligned}
\]
where $d_{\max}\leq\min(D_{\mathrm J},T)-1$.
The resulting signal subspace and noise-subspace projector are
$\mathbf U_{\mathrm s}=\mathbf U_{\mathrm J}(:,1:d_{\mathrm s})$
and
$\boldsymbol{\Pi}_{\mathrm n}
=\mathbf I-\mathbf U_{\mathrm s}\mathbf U_{\mathrm s}^{\mathsf H}$,
respectively.

%\begin{equation}
%d_{\mathrm s}
%=
%\underset{d_{\min}\leq d\leq d_{\max}}
%{\operatorname{arg\,min}}
%%\left\{
%\frac{\beta_{\mathrm{NE}}}{4}
%\left(\frac{T}{D_{\mathrm J}}\right)^2 q_d^2
%+
%2(d+1)
%\right\},
%\label{eq:joint_music_signal_dimension}
%\end{equation}

\subsubsection{JMUSIC Spectrum}

With the multiuser signal and noise subspaces determined, the reduced
delay-AoA candidates are evaluated using a block-subspace JMUSIC
criterion. For candidate $p$, define
$\mathbf C_p=\mathbf U_{\mathrm s}^{\mathsf H}\mathbf B_p
=[\mathbf c_{1,p},\ldots,\mathbf c_{K,p}]$,
where
$\mathbf c_{k,p}=\mathbf U_{\mathrm s,k}^{\mathsf H}\mathbf q_{k,p}$ and
$\mathbf U_{\mathrm s,k}
=\mathbf U_{\mathrm s}(\delta_{k-1}+1:\delta_k,:)$, with
$\delta_k=\sum_{i=1}^{k}D_i$ and $\delta_0=0$.
The quantity $\eta_k(p)=\|\mathbf c_{k,p}\|_2^2$ measures the
UE-$k$ support for candidate $p$.

Define the multiuser support matrix
$\mathbf G_p=\mathbf C_p^{\mathsf H}\mathbf C_p$.
Its diagonal entries $[\mathbf G_p]_{k,k}=\eta_k(p)$ represent
UE-specific support, while the off-diagonal entries
$[\mathbf G_p]_{i,j}=\mathbf c_{i,p}^{\mathsf H}\mathbf c_{j,p}$
capture cross-UE evidence.
Motivated by the matrix-valued subspace criterion in
\cite{Pesavento2002RARE}, each candidate block is evaluated through the
smallest eigenvalue of
$\mathbf B_p^{\mathsf H}\boldsymbol{\Pi}_{\mathrm n}\mathbf B_p$.

\begin{proposition}
\label{prop:joint_music_support}
For candidate $p$, let
$\mu_p\triangleq\lambda_{\max}(\mathbf G_p)$. The proposed JMUSIC
spectrum can be written as
\begin{equation}
P_{\mathrm J}(p)
=
\frac{1}{
\lambda_{\min}
\left(
\mathbf B_p^{\mathsf H}
\boldsymbol{\Pi}_{\mathrm n}
\mathbf B_p
\right)}
=
\frac{1}{1-\mu_p}.
\label{eq:joint_music_spectrum}
\end{equation}
Moreover, the joint support satisfies
\begin{equation}
\mu_p
\geq
\max_k \eta_k(p).
\label{eq:joint_music_support_lower_bound}
\end{equation}
\end{proposition}

The proof is provided in Appendix~\ref{app:joint_music_proof}.
Proposition~\ref{prop:joint_music_support} shows that the joint criterion
preserves private candidates supported by an individual UE, while the
off-diagonal entries of $\mathbf G_p$ allow cross-UE evidence to further
enhance the support of shared propagation components.

\subsubsection{Global-Pool Selection}

The JMUSIC spectrum ranks the reduced delay-AoA candidates according
to their consistency with the joint signal subspace. A global pool encompassing the delay-AoA components of all UEs
must then be extracted from this ranking while suppressing redundant
neighboring candidates and determining the number of retained components.

We first apply two-dimensional non-maximum suppression (NMS) in descending
order of $P_{\mathrm J}(p)$. Specifically, a lower-ranked candidate $p$
is suppressed if there exists a previously retained candidate $p'$ such that
$|\bar\tau_p-\bar\tau_{p'}|\leq\Delta_{\tau}^{\mathrm{NMS}}$ and
$|\bar\theta_p-\bar\theta_{p'}|\leq\Delta_{\theta}^{\mathrm{NMS}}$.
The surviving candidates are denoted by
$\{\widetilde p_q\}_{q=1}^{Q}$ and ordered such that
$P_{\mathrm J}(\widetilde p_1)\geq\cdots\geq
P_{\mathrm J}(\widetilde p_Q)$.

Each surviving candidate is then examined in this fixed order to determine
whether it provides sufficient additional explanatory power for $\widetilde{\mathbf Y}_{\mathrm J}$. For any trial support
$\mathcal S\subseteq\{\widetilde p_1,\ldots,\widetilde p_Q\}$,
define the corresponding block dictionary as
$\mathbf B_{\mathcal S}=[\mathbf B_p]_{p\in\mathcal S}$.
Its coefficient matrix is estimated by regularized least squares as
$\widehat{\mathbf X}_{\mathcal S}
=
(\mathbf B_{\mathcal S}^{\mathsf H}\mathbf B_{\mathcal S}
+\lambda_{\mathrm J}\mathbf I)^{-1}
\mathbf B_{\mathcal S}^{\mathsf H}
\widetilde{\mathbf Y}_{\mathrm J}$,
where $\lambda_{\mathrm J}>0$ is a regularization coefficient. The corresponding residual sum of squares (RSS) is
$\operatorname{RSS}(\mathcal S)
=
\|\widetilde{\mathbf Y}_{\mathrm J}
-\mathbf B_{\mathcal S}\widehat{\mathbf X}_{\mathcal S}\|_{\mathrm F}^{2}$.

The accepted support is initialized as
$\mathcal S_0=\varnothing$, with
$\operatorname{RSS}(\mathcal S_0)=\|\widetilde{\mathbf Y}_{\mathrm J}\|_{\mathrm F}^{2}$.
For candidate $\widetilde p_q$, let $\mathcal S_L$ denote the support
formed by the $L$ previously accepted candidates. Its incremental
contribution is measured relative to the initial residual energy as
\begin{equation}
g_{\mathrm J}(q)
=
\frac{
\operatorname{RSS}(\mathcal S_L)
-
\operatorname{RSS}
\left(
\mathcal S_L\cup\{\widetilde p_q\}
\right)
}{
\operatorname{RSS}(\mathcal S_0)
}.
\label{eq:joint_music_global_rss_reduction}
\end{equation}
If $g_{\mathrm J}(q)\geq\epsilon_{\mathrm J}^{\mathrm{RSS}}$,
the candidate is accepted and the support is updated as
$\mathcal S_{L+1}=\mathcal S_L\cup\{\widetilde p_q\}$;
otherwise, it is discarded and the scan proceeds to the next candidate $\widetilde p_{q+1}$.
Here, $\epsilon_{\mathrm J}^{\mathrm{RSS}}>0$ denotes the redundancy-screening threshold.

For each accepted model order $L$, the corresponding minimum description
length (MDL) criterion \cite{ref_mdl_wax} is recorded as
\begin{equation}
\operatorname{MDL}_{\mathrm J}(L)
\!=\!
D_{\mathrm J}T
\log\!\frac{\operatorname{RSS}(\mathcal S_L)}
{D_{\mathrm J}T}
\!+\!
L(1\!+\!KT)\log(D_{\mathrm J}T),
\label{eq:joint_music_global_mdl}
\end{equation}
After all $Q$ candidates have been scanned, let
$\{\mathcal S_L\}_{L=0}^{Q_1}$ denote the resulting nested sequence of
accepted supports, where $Q_1$ is the total number of RSS-accepted
candidates.

The final global model order is selected from the recorded MDL sequence.
Let
$\operatorname{MDL}_{\mathrm J,\min}
=
\min_{1\leq L\leq Q_1}
\operatorname{MDL}_{\mathrm J}(L)$.
The selected order is
\[
\begin{aligned}
\widehat L_{\mathrm J}
&\!=\!\max\Biggl\{1\leq L\leq Q_1:
\frac{
\operatorname{MDL}_{\mathrm J}(L)
\!-\!\operatorname{MDL}_{\mathrm J,\min}
}{
|\operatorname{MDL}_{\mathrm J,\min}|
}
\leq\epsilon_{\mathrm J}^{\mathrm{MDL}}
\Biggr\},
\end{aligned}
\]
where $\epsilon_{\mathrm J}^{\mathrm{MDL}}>0$ allows a small deviation from the minimum MDL to avoid over-pruning.
The resulting global delay-AoA pool is
$\mathcal P_{\mathrm J}
=
\{(\bar\tau_p,\bar\theta_p):
p\in\mathcal S_{\widehat L_{\mathrm J}}\}$.

\subsubsection{Global-Pool Refinement} Since the selected global components are still restricted to the discrete delay-AoA grid, they need to be refined from $\widetilde{\mathbf Y}_{\mathrm J}$ before
UE-specific association. We use a Gauss-Seidel (GS)  component-cancellation update over the
multiuser candidate blocks defined in
\eqref{eq:joint_music_block_steering} \cite{ref_efficient_async}.

For each selected candidate
$p\in\mathcal S_{\widehat L_{\mathrm J}}$, the contributions of the other selected candidates are removed to form
the path-isolated observation $\mathbf R_{\mathrm J,p}
=
\widetilde{\mathbf Y}_{\mathrm J}
-
\sum_{p'\in\mathcal S_{\widehat L_{\mathrm J}}\setminus\{p\}}
\mathbf B_{p'}\widehat{\mathbf X}_{p'} ,$
where $\widehat{\mathbf X}_{p'}$ is the corresponding coefficient
block from the regularized-LS solution over
$\mathcal S_{\widehat L_{\mathrm J}}$.

The delay and AoA of candidate $p$ are then refined over local grids
centered at $(\bar\tau_p,\bar\theta_p)$, defined as
$\mathcal G_{\tau,p}
=\{\bar\tau_p+i\Delta_\tau/4:i=-4,\ldots,4\}$
and
$\mathcal G_{\theta,p}
=\{\bar\theta_p+j\Delta_\theta/4:j=-2,\ldots,2\}$.
For each trial pair $(\tau,\theta)$, the UE-specific responses forming
$\mathbf B_p$ are re-evaluated according to
\eqref{eq:joint_music_block_steering}.
The updated parameters are obtained as
\begin{equation}
(\widehat{\tau}_p^{\mathrm{new}},
\widehat{\theta}_p^{\mathrm{new}})
=
\arg\max_{(\tau,\theta)\in
\mathcal G_{\tau,p}\times\mathcal G_{\theta,p}}
\|\mathbf B_p^{\mathsf H}\mathbf R_{\mathrm J,p}\|_{\mathrm F}^2 .
\label{eq:joint_music_global_refine}
\end{equation}

After one GS sweep over $\mathcal S_{\widehat L_{\mathrm J}}$
, the coefficient matrix
$\widehat{\mathbf X}_{\mathrm J}$ is jointly re-estimated by regularized
LS using the refined delay-AoA parameters. The refined global pool is
$\widehat{\mathcal P}_{\mathrm J}
=\{(\widehat{\tau}_p^{\mathrm{new}},
\widehat{\theta}_p^{\mathrm{new}}):
p\in\mathcal S_{\widehat L_{\mathrm J}}\}$.

The refined global pool, termed JMUSIC-GLOBAL,  can directly serve as a scene-level sensing
output, e.g., for identifying dominant reflectors or scattering regions
in the delay-AoA domain and monitoring common environmental components.
When UE-specific information is required, the refined global components
are further associated with individual UEs as described next.

\subsection{UE-Specific Association and Residual-TO Refinement}
\label{subsec:joint_music_per_user_association}

The refined global pool provides a scene-level set of delay-AoA components,
but does not specify which components are present in the observation of each
UE. We therefore alternate UE-specific concentrated DML association with
residual-TO refinement over the fixed refined global pool
$\widehat{\mathcal P}_{\mathrm J}$ for $I_{\mathrm{TO,max}}$ iterations.
At each iteration, the current UE-specific delay-AoA support
obtained by DML association provides the signal reconstruction
for residual-TO refinement, while the residual-TO-compensated
observations are used to re-estimate
the UE-specific support for the next refinement iteration. 

\subsubsection{DML Association and Model-Order Selection}

For UE $k$, the $\widehat L_{\mathrm J}$ components in the refined global
pool are sequentially examined to determine its UE-specific delay-AoA
support. Hereafter, $\mathbf q_{k,\ell}$ denotes the normalized response of
UE $k$ corresponding to the $\ell$th refined global component, while
$\mathbf q_{k,\mathrm{ref}}\triangleq
\widehat{\boldsymbol{\phi}}_{k,\mathrm{ref}}/
\|\widehat{\boldsymbol{\phi}}_{k,\mathrm{ref}}\|_2$
denotes the normalized LoS reference atom. Starting from
$\mathcal S_{k,0}=\varnothing$, a nested support sequence is constructed by
concentrated DML. For a trial support $\mathcal S$, define
$\mathbf Q_{k,\mathcal S}
=[\mathbf q_{k,\mathrm{ref}},
[\mathbf q_{k,\ell}]_{\ell\in\mathcal S}]$.
For any observation matrix $\mathbf Z$, define the regularized LS
coefficient estimate
$\widehat{\mathbf X}_k(\mathcal S;\mathbf Z)
=
(\mathbf Q_{k,\mathcal S}^{\mathsf H}
\mathbf Q_{k,\mathcal S}
+\lambda_{\mathrm U}\mathbf I)^{-1}
\mathbf Q_{k,\mathcal S}^{\mathsf H}\mathbf Z$
and the corresponding
$\operatorname{RSS}_k(\mathcal S;\mathbf Z)
=
\|\mathbf Z
-\mathbf Q_{k,\mathcal S}
\widehat{\mathbf X}_k(\mathcal S;\mathbf Z)\|_{\mathrm F}^{2}$,
where $\lambda_{\mathrm U}>0$ is the UE-specific regularization
coefficient. The initial DML association uses
$\mathbf Z=\widetilde{\mathbf Y}_{k}^{\mathrm v}$.

At model order $L$, select the next UE-$k$ candidate by
\begin{equation}
\widehat{\ell}_{k,L}
=
\underset{
\ell\in
\{1,\ldots,\widehat L_{\mathrm J}\}
\setminus
\mathcal S_{k,L-1}
}{
\operatorname{arg\,min}}
\;
\operatorname{RSS}_k
\left(
\mathcal S_{k,L-1}\cup\{\ell\};
\mathbf Z
\right),
\label{eq:joint_music_user_selection}
\end{equation}
and the support is updated as
$\mathcal S_{k,L}
=\mathcal S_{k,L-1}\cup\{\widehat{\ell}_{k,L}\}$.
The corresponding MDL criterion is recorded as
\begin{equation}
\operatorname{MDL}_k(L)
\!\!=\!\!
D_kT\!
\log\!\frac{\operatorname{RSS}_k(\mathcal S_{k,L};\!\mathbf Z)}
{D_kT}
\!\!+\!\!
L(1\!+\!T)\!\log(D_kT).
\label{eq:joint_music_user_mdl}
\end{equation}
Repeating the above selection for up to
$\widehat L_{\mathrm J}$ iterations
produces the nested support sequence
$\{\mathcal S_{k,L}\}_{L=0}^{\widehat L_{\mathrm J}}$
and the corresponding MDL values.

The UE-specific model order is then determined from the recorded MDL
sequence. Let
$\operatorname{MDL}_{k,\min}
=
\min_{1\leq L\leq\widehat L_{\mathrm J}}
\operatorname{MDL}_k(L)$.
The selected model order is
\[
\begin{aligned}
\widehat L_{k,\mathrm N}
&\!=\!\max\Biggl\{1\!\leq\! L\leq\widehat L_{\mathrm J}:
\frac{
\operatorname{MDL}_k(L)
\!-\!\operatorname{MDL}_{k,\min}
}{
|\operatorname{MDL}_{k,\min}|
}
\!\leq\!\epsilon_{\mathrm U}^{\mathrm{MDL}}
\Biggr\},
\end{aligned}
\]
%The resulting UE-specific delay-AoA support is $\mathcal P_k = \{(\widehat{\tau}_{k,\ell}, \widehat{\theta}_{k,\ell})\}_{\ell=1}^{\widehat L_{k,\mathrm N}}$. 
where $\epsilon_{\mathrm U}^{\mathrm{MDL}}>0$ is the relative MDL
tolerance. The initial DML association yields the model order
$\widehat L_{k,\mathrm N}$ and support
$\mathcal S_k^{(0)}\triangleq\mathcal S_{k,\widehat L_{k,\mathrm N}}$
for subsequent residual-TO refinement.

\subsubsection{DML-Aided Residual-TO Refinement}

The residual-TO update in Section~III-C requires
$\widehat{\mathbf s}_k^{(i)}[t]$ and $\kappa_k^{(i)}$, which are obtained
from the current UE-specific DML reconstruction. We initialize
$\widetilde{\mathbf Y}_{k,\mathrm{RTO}}^{\mathrm v(0)}
=\widetilde{\mathbf Y}_k^{\mathrm v}$ and
$\widehat{\epsilon}_{k,\mathrm{RTO}}^{(0)}[t]=0$.

For residual-TO refinement iteration $i$, let
$\mathcal S_k^{(i-1)}$ denote the current UE-specific support.
The reconstructed signal is then obtained using the regularized LS
estimator defined above as
$\widehat{\mathbf Y}_k^{\mathrm v(i)}
=
\mathbf Q_{k,\mathcal S_k^{(i-1)}}
\widehat{\mathbf X}_k
(\mathcal S_k^{(i-1)};
\widetilde{\mathbf Y}_{k,\mathrm{RTO}}^{\mathrm v(i-1)})$.
Let $\widehat{\mathbf y}_k^{(i)}[t]$ denote its $t$th column; then
$\widehat{\mathbf s}_k^{(i)}[t]
=
\widehat{\mathbf y}_k^{(i)}[t]$.

Unlike Group-SBL, JMUSIC does not provide a posterior estimate of
$\kappa_k^{(i)}$. We therefore estimate it from the DML reconstruction
error
$\mathbf R_k^{(i)}
\triangleq
\widetilde{\mathbf Y}_{k,\mathrm{RTO}}^{\mathrm v(i-1)}
-
\widehat{\mathbf Y}_k^{\mathrm v(i)}$.
Assuming the entries of $\mathbf R_k^{(i)}$ are i.i.d.
$\mathcal{CN}(0,(\kappa_k^{(i)})^{-1})$, maximizing the corresponding
likelihood gives
$\kappa_k^{(i)}=D_kT/\|\mathbf R_k^{(i)}\|_{\mathrm F}^{2}$.

The resulting $\widehat{\mathbf s}_k^{(i)}[t]$ and
$\kappa_k^{(i)}$ are used in the common residual-TO update of
Section~\ref{subsec:residual_sync} to obtain
$\widetilde{\mathbf Y}_{k,\mathrm{RTO}}^{\mathrm v(i)}$. After $I_{\mathrm{TO,max}}$ iterations, the final residual-TO-refined
observation is denoted by
$\widetilde{\mathbf Y}_{k}^{\mathrm v\star}
\triangleq
\widetilde{\mathbf Y}_{k,\mathrm{RTO}}^{\mathrm v(I_{\mathrm{TO,max}})}$.
The complete alternating procedure is summarized in
Algorithm~\ref{alg:joint_music}.

\begin{algorithm}[t]
\caption{ JMUSIC Procedure}
\label{alg:joint_music}
\begin{algorithmic}[1]
\REQUIRE
$\{\widetilde{\mathbf Y}_k^{\mathrm v},
\boldsymbol{\Phi}_{k,\Omega}\}_{k=1}^{K}$, and $I_{\mathrm{TO,max}}$.

\STATE Construct the stacked multiuser observation
$\widetilde{\mathbf Y}_{\mathrm J}$ via
\eqref{eq:joint_music_stacked_observation}, and obtain
$d_{\mathrm s}$, $\mathbf U_{\mathrm s}$, and
$\boldsymbol{\Pi}_{\mathrm n}$ from its compact SVD.

\STATE Evaluate
$\{P_{\mathrm J}(p)\}_{p=1}^{P_\Omega}$
via \eqref{eq:joint_music_spectrum}, determine
$\widehat L_{\mathrm J}$ using the MDL criterion in
\eqref{eq:joint_music_global_mdl} to construct the global delay-AoA pool, and refine it via
\eqref{eq:joint_music_global_refine} to obtain
$\widehat{\mathcal P}_{\mathrm J}$. 

\FOR{$k=1,\ldots,K$}

\STATE Obtain the initial UE-specific support
$\mathcal S_k^{(0)}=\mathcal S_{k,\widehat L_{k,\mathrm N}^{(0)}}$
via the DML association and model-order selection in
\eqref{eq:joint_music_user_selection}-\eqref{eq:joint_music_user_mdl};
initialize
$\widetilde{\mathbf Y}_{k,\mathrm{RTO}}^{\mathrm v(0)}
=
\widetilde{\mathbf Y}_k^{\mathrm v}$
and
$\widehat{\epsilon}_{k,\mathrm{RTO}}^{(0)}[t]=0$.

\FOR{$i=1,\ldots,I_{\mathrm{TO,max}}$}

\STATE Form
$\widehat{\mathbf Y}_k^{\mathrm v(i)}
=
\mathbf Q_{k,\mathcal S_k^{(i-1)}}
\widehat{\mathbf X}_k
(\mathcal S_k^{(i-1)};
\widetilde{\mathbf Y}_{k,\mathrm{RTO}}^{\mathrm v(i-1)})$;
compute
$\kappa_k^{(i)}=D_kT/\|\mathbf R_k^{(i)}\|_{\mathrm F}^{2}$.
Apply
\eqref{eq:residual_to_update}-\eqref{eq:residual_to_accumulation}
to update the residual TO and form
$\widetilde{\mathbf Y}_{k,\mathrm{RTO}}^{\mathrm v(i)}$
using \eqref{eq:centered_rto_compensation}.

\STATE Repeat the UE-specific DML association and MDL evaluation with
$\mathbf Z=
\widetilde{\mathbf Y}_{k,\mathrm{RTO}}^{\mathrm v(i)}$
over the fixed $\widehat{\mathcal P}_{\mathrm J}$ in
\eqref{eq:joint_music_user_selection}-\eqref{eq:joint_music_user_mdl},
obtain $\widehat L_{k,\mathrm N}^{(i)}$, and set
$\mathcal S_k^{(i)}
=
\mathcal S_{k,\widehat L_{k,\mathrm N}^{(i)}}$.

\ENDFOR
\STATE Update the final UE-specific delay-AoA support with the LoS reference as
$\mathcal P_k
\!=\!
\{(\widehat{\tau}_{k,\ell},\widehat{\theta}_{k,\ell}):
\ell\in\mathcal S_k^{(I_{\mathrm{TO,max}})}\}
\cup
\{(0,\widehat{\theta}_{k,\mathrm{ref}})\}$, set $\widehat L_k=|\mathcal P_k|$. 

\ENDFOR

\textbf{Output:} $\widehat{\mathcal P}_{\mathrm J}$,
$\{\widetilde{\mathbf Y}_{k}^{\mathrm v\star}\}_{k=1}^{K}$,
and
$\{\mathcal P_k\}_{k=1}^{K}$.

\end{algorithmic}
\end{algorithm}

\section{Delay-AoA Refinement and Doppler Estimation}

As illustrated in Fig.~\ref{fig:overall_framework}, the UE-specific delay-AoA support estimates obtained
from either multiuser sensing estimator are
subsequently processed by the same procedure comprising delay-AoA refinement, residual
common-phase compensation, and one-dimensional Doppler estimation.

\subsection{Local-grid Delay-AoA Refinement}
For either multiuser estimator, let
$\widetilde{\mathbf Y}_{k}^{\mathrm v\star}$
denote the final residual-TO-refined observation of UE $k$.  We first apply GS refinement to the UE-specific delay-AoA estimates. To suppress inter-path leakage, each
path is sequentially refined after cancelling the estimated contributions
of the others.

For UE \(k\), let
\(\{(\widehat{\tau}_{k,\ell}^{(r)},
\widehat{\theta}_{k,\ell}^{(r)})\}_{\ell=1}^{\widehat L_k}\)
denote the estimates at refinement sweep \(r\). Define
\(\widehat{\boldsymbol{\phi}}_{k,\ell}^{(r)}
=\mathbf d_k(\widehat{\tau}_{k,\ell}^{(r)})
\otimes\mathbf a(\widehat{\theta}_{k,\ell}^{(r)})\)
and
\(\widehat{\boldsymbol{\Phi}}_k^{(r)}
=[\widehat{\boldsymbol{\phi}}_{k,1}^{(r)},\ldots,
\widehat{\boldsymbol{\phi}}_{k,\widehat L_k}^{(r)}]\).
The path coefficients are jointly updated by regularized least squares as
$\widehat{\mathbf C}_k^{(r)}
=
[(\widehat{\boldsymbol{\Phi}}_k^{(r)})^{\mathsf H}
\widehat{\boldsymbol{\Phi}}_k^{(r)}
+\lambda_{\mathrm{LS}}\mathbf I]^{-1}
(\widehat{\boldsymbol{\Phi}}_k^{(r)})^{\mathsf H}
\widetilde{\mathbf Y}_{k}^{\mathrm v\star}$,
where row $\widehat{\mathbf c}_{k,\ell}^{(r)}$ contains the
$T$-snapshot coefficient sequence of path $\ell$.

For path \(\ell\), canceling the other paths gives
\begin{equation}
\mathbf R_{k,\ell}^{(r)}
=
\widetilde{\mathbf Y}_{k}^{\mathrm v\star}
-
\widehat{\boldsymbol{\Phi}}_k^{(r)}
\widehat{\mathbf C}_k^{(r)}
+
\widehat{\boldsymbol{\phi}}_{k,\ell}^{(r)}
\widehat{\mathbf c}_{k,\ell}^{(r)}.
\label{eq:path_specific_residual}
\end{equation}
Define the local matching score as
\(\mathcal M_{k,\ell}^{(r)}(\tau,\theta)
\triangleq
\|\boldsymbol{\phi}_k^{\mathsf H}(\tau,\theta)
\mathbf R_{k,\ell}^{(r)}\|_2^2/
\|\boldsymbol{\phi}_k(\tau,\theta)\|_2^2\).
Using the local-grid construction in
\eqref{eq:joint_music_global_refine}, path $\ell$ is refined around
$(\widehat{\tau}_{k,\ell}^{(r)},
\widehat{\theta}_{k,\ell}^{(r)})$ as
\begin{equation}
(\widehat{\tau}_{k,\ell}^{\mathrm{new}},
\widehat{\theta}_{k,\ell}^{\mathrm{new}})
=
\arg\max_{(\tau,\theta)\in
\mathcal G_{\tau,k,\ell}^{(r)}
\times
\mathcal G_{\theta,k,\ell}^{(r)}}
\mathcal M_{k,\ell}^{(r)}(\tau,\theta).
\end{equation}
For the candidate pair, define
$\widehat{\boldsymbol{\phi}}_{k,\ell}^{\mathrm{new}}
=
\boldsymbol{\phi}_k(
\widehat{\tau}_{k,\ell}^{\mathrm{new}},
\widehat{\theta}_{k,\ell}^{\mathrm{new}})$
and update its coefficient sequence as $\widehat{\mathbf c}_{k,\ell}^{\mathrm{new}}
=
\frac{
(\widehat{\boldsymbol{\phi}}_{k,\ell}^{\mathrm{new}})^{\mathsf H}
\mathbf R_{k,\ell}^{(r)}
}{
\|\widehat{\boldsymbol{\phi}}_{k,\ell}^{\mathrm{new}}\|_2^2
}$.
If the candidate pair improves
$\mathcal M_{k,\ell}^{(r)}$, it replaces the current path estimate together with
$\widehat{\boldsymbol{\phi}}_{k,\ell}^{\mathrm{new}}$ and
$\widehat{\mathbf c}_{k,\ell}^{\mathrm{new}}$; otherwise, the current estimate is retained.
Thus, subsequent paths are processed using the latest available estimates within the same sweep.

After all $\widehat L_k$ paths are processed, the resulting atom matrix is denoted by $\widehat{\boldsymbol{\Phi}}_k^{(r+1)}$, and the path coefficients $\widehat{\mathbf C}_k^{(r+1)}$ are jointly re-estimated by the same regularized-LS update. Define
$\mathcal Q_k^{(r)}
=
\|\widetilde{\mathbf Y}_{k}^{\mathrm v\star}
-\widehat{\boldsymbol{\Phi}}_k^{(r)}
\widehat{\mathbf C}_k^{(r)}\|_{\mathrm F}^{2}$.
The refinement terminates when
$(\mathcal Q_k^{(r)}-\mathcal Q_k^{(r+1)})/\mathcal Q_k^{(r)}
<\varepsilon_{\mathrm{GS}}$
or when the maximum number of refinement sweeps $I_r^{\max}$ is reached, yielding the refined delay-AoA estimates
$\{(\widehat{\tau}_{k,\ell}^{\star},
\widehat{\theta}_{k,\ell}^{\star}\}_{\ell=1}^{\widehat L_k}$.

\subsection{Residual Common-Phase Compensation}
After delay-AoA refinement, define
$\widehat{\boldsymbol{\phi}}_{k,\ell}
=
\boldsymbol{\phi}_k(
\widehat{\tau}_{k,\ell}^{\star},
\widehat{\theta}_{k,\ell}^{\star})$.
%The observations used for Doppler estimation have already undergone the residual-TO refinement described in Section~III-C. let
Let $\widetilde{\mathbf y}_{k}^{\star}[t]$
denote the $t$th column of $\widetilde{\mathbf Y}_{k}^{\mathrm v\star}$.
We now estimate the residual common phase $\xi_k[t]$ defined in
Section~\ref{subsec:residual_sync}, which remains after the
LoS-referenced frontend calibration and is absorbed into the complex
coefficients during delay-AoA estimation. The observation can therefore
be written as
\begin{equation}
\widetilde{\mathbf y}_{k}^{\star}[t]
\approx
e^{j\xi_k[t]}
\sum_{\ell=1}^{\widehat L_k}
\alpha_{k,\ell}
e^{j2\pi\nu_{k,\ell}t_t}
\widehat{\boldsymbol{\phi}}_{k,\ell}
+
\widetilde{\mathbf e}_k[t].
\label{eq:residual_cp_doppler_model}
\end{equation}

To suppress interference from the other retained paths, let
$\widehat{\boldsymbol{\Phi}}_{k,-\ell}$ collect all refined atoms except
$\widehat{\boldsymbol{\phi}}_{k,\ell}$, and define the regularized
nuisance-suppression operator $\mathbf P_{k,-\ell}^{\perp}
=
\mathbf I
-
\widehat{\boldsymbol{\Phi}}_{k,-\ell}
(
\widehat{\boldsymbol{\Phi}}_{k,-\ell}^{\mathsf H}
\widehat{\boldsymbol{\Phi}}_{k,-\ell}
+
\lambda_{\mathrm{LS}}\mathbf I
)^{-1}
\widehat{\boldsymbol{\Phi}}_{k,-\ell}^{\mathsf H}.$
For simplicity, define
$\mathbf u_{k,\ell}
\triangleq
\mathbf P_{k,-\ell}^{\perp}
\widehat{\boldsymbol{\phi}}_{k,\ell}$.
After LoS-referenced calibration and the subsequent residual-TO refinement, the quasi-static reference path is located
at zero relative delay and is therefore identified by
$r_k=\arg\min_{\ell}|\widehat{\tau}_{k,\ell}^{\star}|$.
Suppressing the other retained paths yields the \begin{equation}
z_{k,r_k}[t]
\!=\!
\frac{
\mathbf u_{k,r_k}^{\mathsf H}
\mathbf P_{k,-r_k}^{\perp}
\widetilde{\mathbf y}_{k}^{\star}[t]
}{
\|\mathbf u_{k,r_k}\|_2^2
}
\!\approx\!
\alpha_{k,r_k}e^{j\xi_k[t]}
\!+\!\widetilde e_{k,r_k}[t],
\label{eq:reference_common_phase}
\end{equation}
Since the reference path is quasi-static, its Doppler is negligible and
$\alpha_{k,r_k}$ is approximately snapshot invariant, yielding
$z_{k,r_k}[t]\approx\alpha_{k,r_k}e^{j\xi_k[t]}$.
Using snapshot~1 as the phase reference obtains
$z_{k,r_k}[t]z_{k,r_k}^{*}[1]
\approx|\alpha_{k,r_k}|^2e^{j(\xi_k[t]-\xi_k[1])}$,
which eliminates the unknown constant phase of the reference-path gain.
We therefore estimate the residual common phase by
\begin{equation}
\widehat{\xi}_k[t]
=
\angle
\left\{
z_{k,r_k}[t]z_{k,r_k}^{*}[1]
\right\}.
\label{eq:residual_common_phase}
\end{equation}
The corresponding phase correction is then applied as
\(\overline{\mathbf y}_k[t]
=e^{-j\widehat{\xi}_k[t]}
\widetilde{\mathbf y}_{k}^{\star}[t]\).

\subsection{Doppler Estimation}
Conditioned on the refined delay-AoA estimates and the compensated common
phase, Doppler is the only remaining structural parameter. For path $\ell$,
the corresponding slow-time sequence is obtained as
\begin{equation}
z_{k,\ell}[t]
=
{
\mathbf u_{k,\ell}^{\mathsf H}
\mathbf P_{k,-\ell}^{\perp}
\overline{\mathbf y}_k[t]
}/{
\|\mathbf u_{k,\ell}\|_2^2
},
\label{eq:zkl}
\end{equation}
which approximately follows a single complex exponential in slow time with Doppler frequency $\nu_{k,\ell}$. The Doppler estimate is therefore obtained by the one-dimensional discrete Fourier transform (DFT)
\begin{equation}
\widehat{\nu}_{k,\ell}
=
\arg\max_{\nu\in\mathcal G_{\nu}}
\left|
\sum_{t=1}^{T}
z_{k,\ell}[t]e^{-j2\pi\nu t_t}
\right|^2,
\label{eq:one_dimensional_doppler}
\end{equation}
where $\mathcal G_\nu$ denotes the Doppler search grid, with
its range and resolution specified in Section~VIII.
Finally, three-point interpolation around the dominant Doppler bin
\cite{candan2011dft} is applied to obtain
$\widehat{\nu}_{k,\ell}^{\star}$ and reduce the grid-quantization error. The final UE-specific sensing parameters are
$\{(\widehat{\tau}_{k,\ell}^{\star},
\widehat{\theta}_{k,\ell}^{\star},
\widehat{\nu}_{k,\ell}^{\star})\}_{\ell=1}^{\widehat L_k}$,
for $k=1,\ldots,K$.

\section{Computational Complexity}

We compare the dominant computational complexities of our proposed Group-SBL w/ residual-offset refinement (ROR), JMUSIC w/ ROR, Individual-SBL \cite{wipf2004sbl}, and per-UE MUSIC (PERMUSIC) \cite{schmidt1986music}. Individual-SBL and PERMUSIC independently perform per-UE SBL and MUSIC processing, respectively.  The auxiliary scene-level result JMUSIC-GLOBAL characterizes the coverage and estimation quality of the refined global delay-AoA pool before UE-specific association. 

Since offset mitigation and reduced delay-AoA candidate construction
are common to all schemes, the comparison focuses on the dominant
estimator-specific costs after obtaining $\Omega$.
The subsequent UE-level delay-AoA refinement, residual common-phase
compensation, and one-dimensional Doppler estimation are also common
and incur relatively low computational costs, and are therefore excluded.
 %whereas the residual-TO refinement costs are explicitly included. 
For compactness, define $\mathcal C_{\rm G}\triangleq(2P_\Omega)^3+(2P_\Omega)^2T$ and $\mathcal C_{\rm J}\triangleq D_{\rm J}T^2+P_\Omega(D_{\rm J}d_s+d_sK^2+K^3)$. Let $I_{\rm SBL}$ denote the iteration number of Individual-SBL and $d_{s,k}$ the signal-subspace dimension of PERMUSIC. The global-pool refinement and UE-specific DML association operate on the much smaller retained candidate sets and do not affect the dominant complexity order. The resulting complexities are summarized in Table~\ref{tab}.

\begin{table}[t]
\centering
\caption{Dominant computational complexities of the considered schemes.}
\label{tab}
\renewcommand{\arraystretch}{1.2}
\setlength{\tabcolsep}{3pt}
\begin{tabular}{ll}
\toprule
Scheme & Dominant complexity \\
\midrule

Group-SBL w/ ROR
&
$\mathcal{O}\!\left(
KI_{\max}\mathcal C_{\rm G}
+KN_{\rm TO}^{\max}D_kP_\Omega T
\right)$
\\

Individual-SBL
&
$\mathcal{O}\!\left(
KI_{\rm SBL}[P_\Omega^3+P_\Omega^2T]
\right)$
\\

PERMUSIC
&
$\mathcal{O}\!\left(
K[D_kT^2+P_\Omega D_kd_{s,k}]
\right)$
\\

JMUSIC-GLOBAL
&
$\mathcal{O}\!\left(\mathcal C_{\rm J}\right)$
\\

JMUSIC w/ ROR
&
$\mathcal{O}\!\left(
\mathcal C_{\rm J}
+I_{\rm TO,max}\sum_{k=1}^{K}D_k\widehat L_kT
\right)$
\\

\bottomrule
\end{tabular}
\end{table}

For Group-SBL, the term $KI_{\max}\mathcal C_{\rm G}$ is dominated by the iterative posterior updates over the $2P_\Omega$ shared-private coefficient rows, while $KN_{\rm TO}^{\max}D_kP_\Omega T$ accounts for residual-TO refinement. For JMUSIC, $\mathcal C_{\rm J}$ mainly captures joint subspace construction and candidate-wise JMUSIC spectrum evaluation, while the additional term $I_{\rm TO,max}\sum_{k=1}^{K}D_k\widehat L_kT$ accounts for residual-TO refinement over the retained UE-specific supports. Since $P_\Omega\ll G_\tau G_\theta$, the reduced candidate construction substantially lowers the computational burden of both multiuser schemes.

\section{Simulation Results}

Unless otherwise specified, the simulations adopt a 5G
NR-compatible
uplink configuration with carrier frequency
$f_c=3.5$ GHz and channel bandwidth $B_{\mathrm{ch}}=100$ MHz.
The subcarrier spacing is $\Delta f=60$ kHz, corresponding to 135 PRBs
and $N=1620$ active subcarriers, with 12 subcarriers per PRB.
For $K=8$, each UE is allocated 10 contiguous
PRBs, corresponding to $N_k=120$ subcarriers, following localized
PUSCH resource allocation. The BS employs an $M=8$-element
half-wavelength ULA and collects $T=32$ CSI snapshots with
snapshot interval $T_A=0.25$ ms.

The UEs are partitioned into $C_{\mathrm{true}}=3$ propagation groups.
Each UE contains $L_{\mathrm{sh}}=4$ group-shared paths, including three
non-line-of-sight (NLoS) paths and one dominant quasi-static LoS reference, together with
$L_{\mathrm{pr}}=1$ private NLoS path. The path parameters satisfy
$\tau_{k,\ell}\in[0,\tau_{\max}]$, with $\tau_{\max}=2.5~\mu\mathrm{s}$,
$\theta_{k,\ell}\in[-90^\circ,90^\circ]$, and
$\nu_{k,\ell}\in[-350,350]$ Hz.

The snapshot-dependent offsets are independently generated across UEs and
snapshots as
$\delta\tau_k[t]\sim\mathcal U(0,20/B)$,
$\delta f_k[t]\sim\mathcal U(0,150~\mathrm{Hz})$, and
$\beta_k[t]\sim\mathcal N(0,(5^\circ)^2)$.
The maximum candidate group number is set to $C=6$.
The Doppler search uses 2001 uniformly spaced points over
$[-500,500]$ Hz. The algorithmic parameters are set to 
$G_{\theta}^{\mathrm{ref}}=181$,
$C_{\mathrm{ref}}=5$,
$G_{\rho}^{\mathrm{ref}}=128$,
$G_{\theta}^{\mathrm{loc}}=41$,
$\Delta_{\theta}^{\mathrm{loc}}=2^\circ$,
$G_{\theta}=181$,
$G_{\tau}=256$,
$G_{\tau}^{\mathrm c}=48$,
$G_{\theta}^{\mathrm c}=60$,
$\epsilon_{\mathrm c}=10^{-2}$,
$I_{\max}^{\mathrm c}=8$,
$M_{\mathrm{loc}}=4$,
$\lambda_{\epsilon}=(50~\mathrm{ns})^{-2}$,
$\Delta\epsilon_{\max}=2~\mathrm{ns}$,
$\boldsymbol{\alpha}_0=10^{-2}\mathbf 1_C$,
$a_{\gamma0}=b_{\gamma0}=a_{\eta0}=b_{\eta0}
=a_{\zeta0}=b_{\zeta0}=10^{-6}$,
$I_{\max}=100$,
$\varepsilon_{\mathrm{VI}}=10^{-3}$,
$I_{\mathrm{TO}}=30$,
$P_{\mathrm{TO}}=5$,
$N_{\mathrm{TO}}^{\max}=10$,
$\lambda_{\mathrm{LS}}=10^{-3}$,
$d_{\max}=T-1$,
$\Delta_{\tau}^{\mathrm{NMS}}=0.05~\mu\mathrm{s}$,
$\Delta_{\theta}^{\mathrm{NMS}}=8^{\circ}$,
$\lambda_{\mathrm U}=\lambda_{\mathrm J}=10^{-6}$,
$\epsilon_{\mathrm J}^{\mathrm{RSS}}=2\times10^{-3}$,
$\epsilon_{\mathrm J}^{\mathrm{MDL}}
=\epsilon_{\mathrm U}^{\mathrm{MDL}}
=5\times10^{-3}$,
$I_{\mathrm{TO,max}}=5$,
$\varepsilon_{\mathrm{GS}}=10^{-4}$,
and $I_r^{\max}=3$.

Let $(\tau_{k,\ell},\theta_{k,\ell},\nu_{k,\ell})$ and
$(\widehat{\tau}_{k,\ell},\widehat{\theta}_{k,\ell},\widehat{\nu}_{k,\ell})$
denote the true and estimated parameters of a matched NLoS path,
respectively, with $\mathcal M$ denoting the matched NLoS path set.
The delay normalized mean-square error (NMSE) and AoA root-mean-square
error (RMSE) are
$\operatorname{NMSE}(\tau)
=\frac{\sum_{(k,\ell)\in\mathcal M}
|\widehat{\tau}_{k,\ell}-\tau_{k,\ell}|^2}
{\sum_{(k,\ell)\in\mathcal M}|\tau_{k,\ell}|^2}$
and
$\operatorname{RMSE}(\theta)
=\sqrt{\frac{1}{|\mathcal M|}
\sum_{(k,\ell)\in\mathcal M}
|\angle\exp\{j(\widehat{\theta}_{k,\ell}-\theta_{k,\ell})\}|^2}$.
Doppler NMSE is similarly computed over matched moving NLoS paths.
For JMUSIC-GLOBAL, the true scene-level NLoS components are $\{(\tau_q^{\mathrm G},\theta_q^{\mathrm G})\}_{q=1}^{L_{\mathrm G}^{\mathrm{true}}}$, where $L_{\mathrm G}^{\mathrm{true}}
=C_{\mathrm{true}}(L_{\mathrm{sh}}-1)+KL_{\mathrm{pr}}$, each shared component uses the average delay and AoA
of the UEs within its corresponding group,
while private components are retained individually. Let $(\widehat{\tau}^{\rm G}_q,\widehat{\theta}^{\rm G}_q)$ denote the corresponding estimate from the refined global delay-AoA pool $\widehat{\mathcal P}_{\mathrm J}$. The global delay NMSE and AoA RMSE are
$\operatorname{NMSE}_{\rm G}(\tau)
=\frac{\sum_{q=1}^{L_{\rm G}^{\rm true}}
|\widehat{\tau}^{\rm G}_q-\tau^{\rm G}_q|^2}
{\sum_{q=1}^{L_{\rm G}^{\rm true}}|\tau^{\rm G}_q|^2}$
and
$\operatorname{RMSE}_{\rm G}(\theta)
=\sqrt{\frac{1}{L_{\rm G}^{\rm true}}
\sum_{q=1}^{L_{\rm G}^{\rm true}}
\left|\angle\exp\{j(\widehat{\theta}^{\rm G}_q
-\theta^{\rm G}_q)\}\right|^2}$.

% =========================
% Offset-removal performance
% =========================
\begin{figure}[t]
    \centering
    \includegraphics[width=0.9\columnwidth]{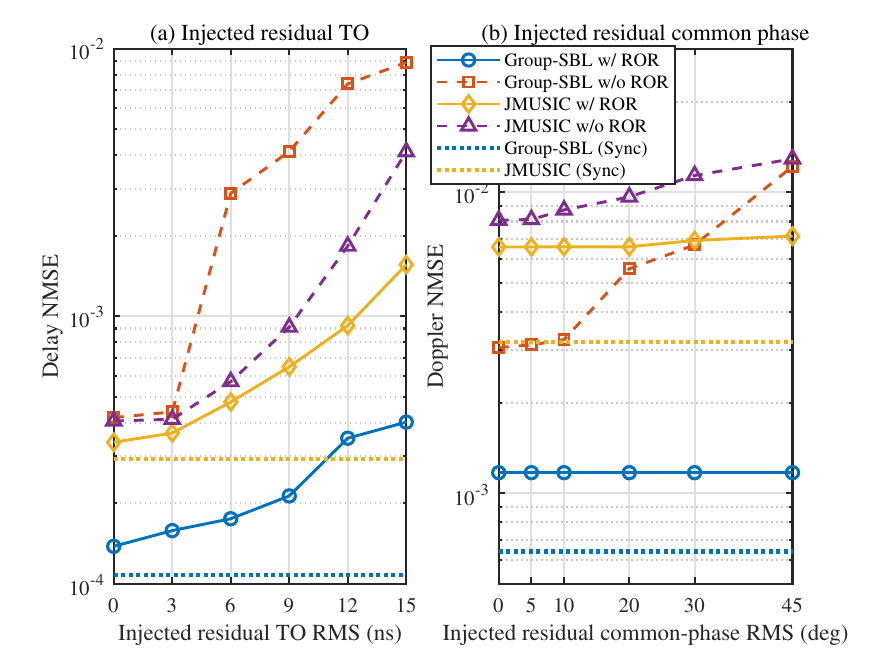}
    \caption{Performance under artificially injected residual synchronization errors after frontend calibration. $\mathrm{SNR}=8$ dB.}
    \label{fig:offset}
\end{figure}

% =========================
% SNR sweep
% =========================
\begin{figure*}[t]
    \centering
    \begin{subfigure}[t]{0.32\textwidth}
        \centering
        \includegraphics[width=\linewidth]{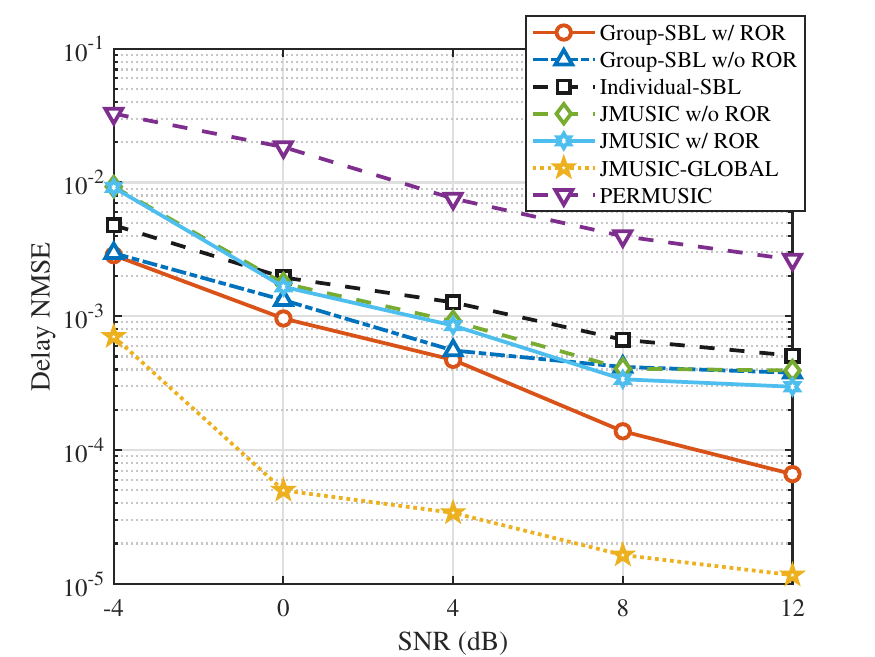}
        \caption{Delay NMSE.}
        \label{fig:snr_delay}
    \end{subfigure}
    \hfill
    \begin{subfigure}[t]{0.32\textwidth}
        \centering
        \includegraphics[width=\linewidth]{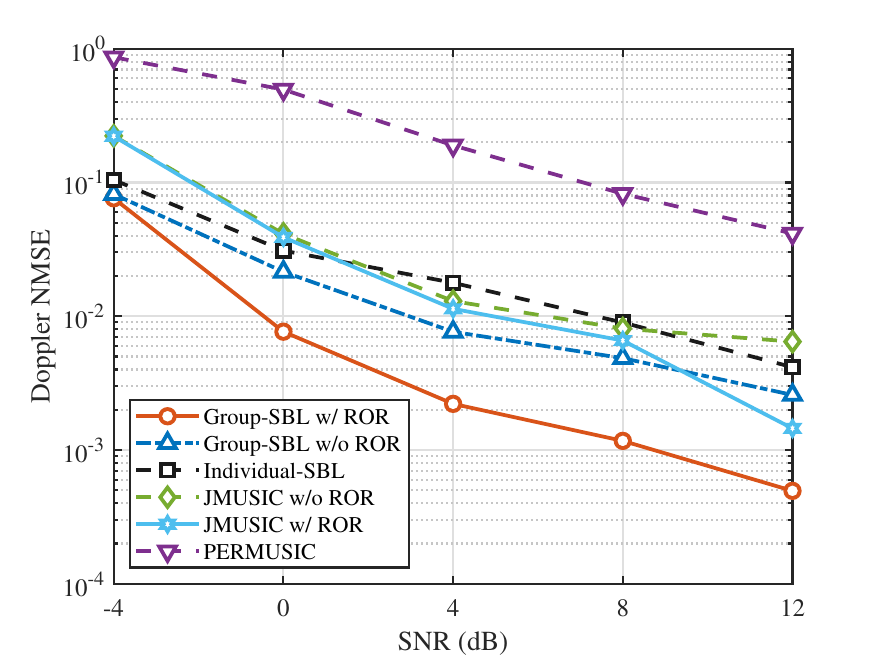}
        \caption{Doppler NMSE.}
        \label{fig:snr_doppler}
    \end{subfigure}
    \hfill
    \begin{subfigure}[t]{0.32\textwidth}
        \centering
        \includegraphics[width=\linewidth]{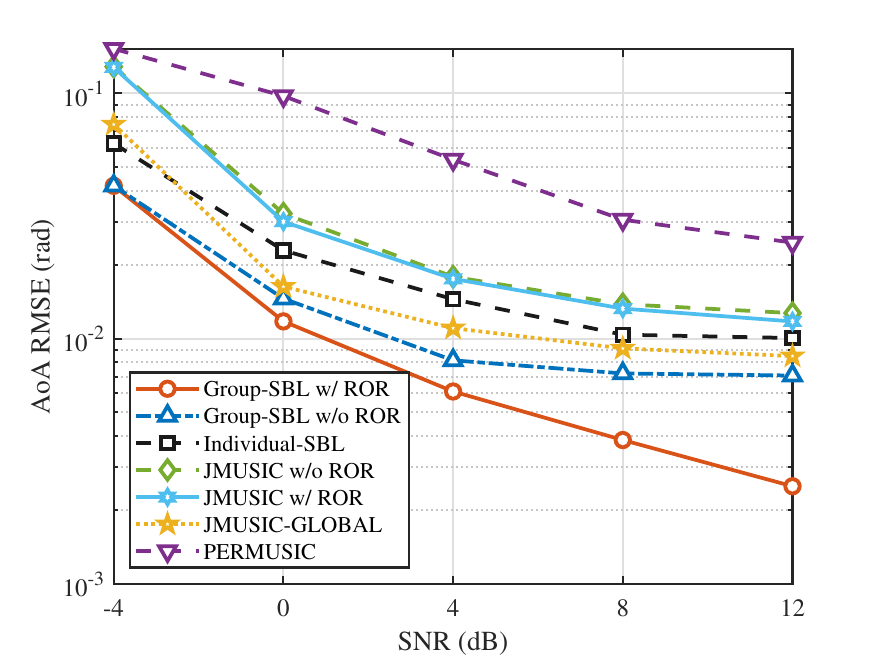}
        \caption{AoA RMSE.}
        \label{fig:snr_aoa}
    \end{subfigure}
    \caption{Delay, Doppler, and AoA estimation performance versus SNR. }
    \label{fig:snr}
\end{figure*}

% =========================
% Nk sweep: (4,1)
% =========================
\begin{figure}[t]
    \centering
    \includegraphics[width=0.9\columnwidth]{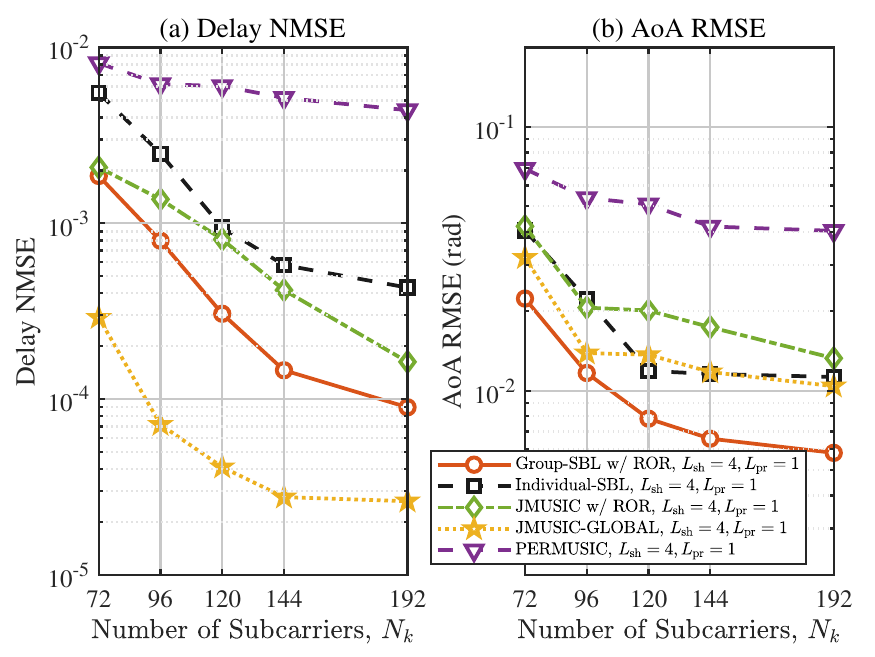}
    \caption{Delay NMSE and AoA RMSE versus $N_k$ for $(L_{\rm sh},L_{\rm pr})=(4,1)$. $T=64$ and $\mathrm{SNR}=4$ dB. }
    \label{fig:nk41}
\end{figure}

% =========================
% Nk sweep: (4,0)
% =========================
\begin{figure}[t]
    \centering
    \includegraphics[width=0.9\columnwidth]{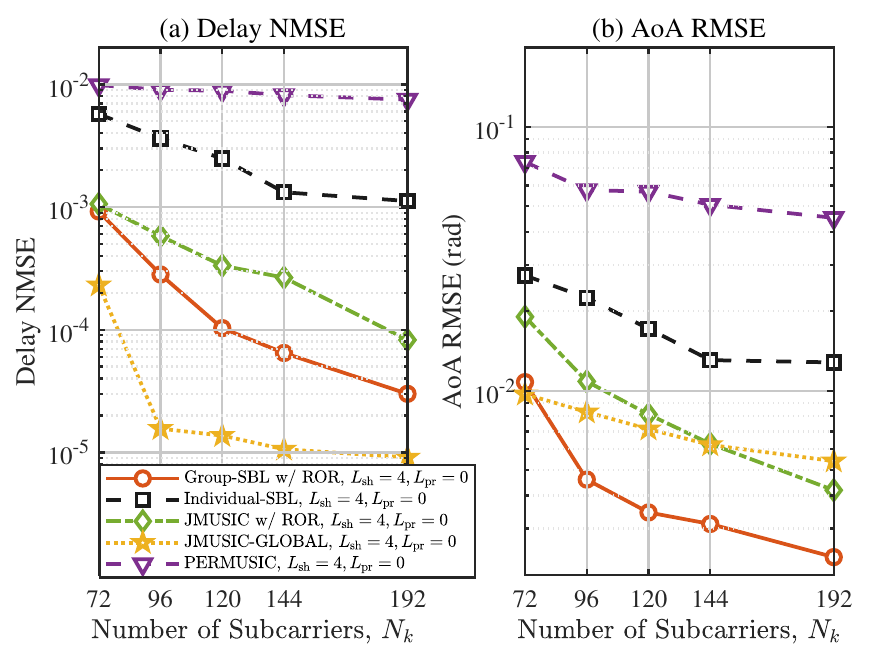}
    \caption{Delay NMSE and AoA RMSE versus $N_k$ for $(L_{\rm sh},L_{\rm pr})=(4,0)$. $T=64$ and $\mathrm{SNR}=4$ dB.}
    \label{fig:nk40}
\end{figure}

% =========================
% T sweep
% =========================
\begin{figure}[t]
    \centering
    \includegraphics[width=0.9\columnwidth]{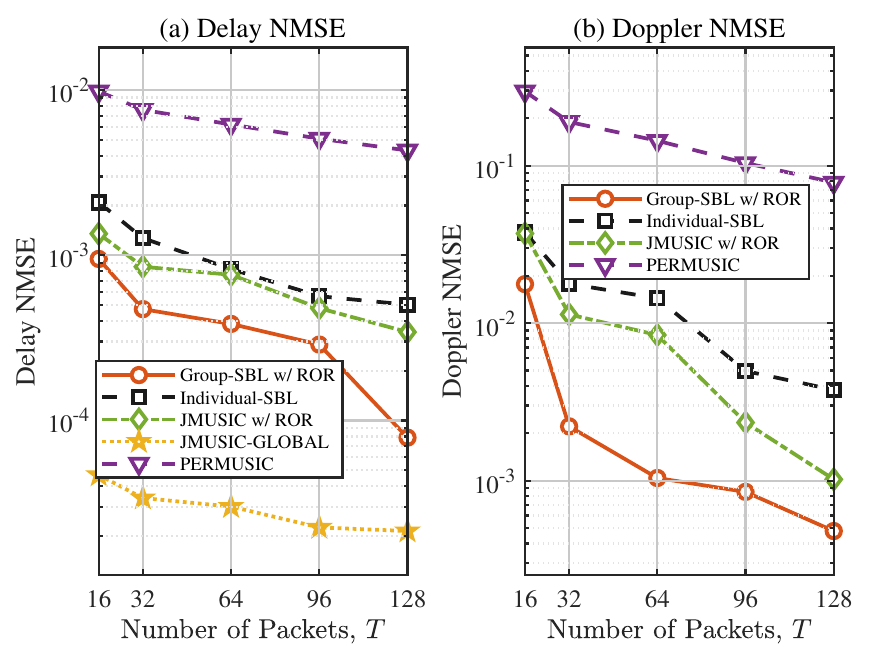}
    \caption{Delay NMSE and Doppler NMSE versus the number of CSI snapshots $T$. $\mathrm{SNR}=4$ dB.}
    \label{fig:T}
\end{figure}

% =========================
% K sweep
% =========================
\begin{figure}[t]
    \centering
    \includegraphics[width=0.9\columnwidth]{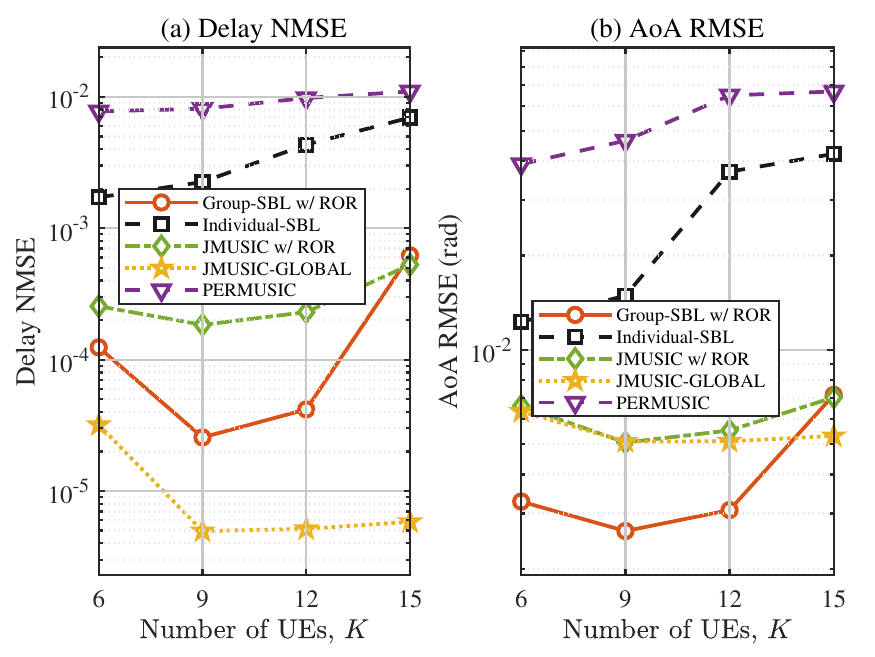}
    \caption{Delay NMSE and AoA RMSE versus the number of UEs $K$ with a fixed total number of subcarriers $N=1620$ for $(L_{\rm sh},L_{\rm pr})=(4,0)$, $T=64$ and $\mathrm{SNR}=4$ dB.}
    \label{fig:K}
\end{figure}

%Fig.~\ref{fig:offset} shows the reference-delay NMSE, reference-AoA RMSE, and residual CFO/PO phase RMSE versus SNR. All three metrics decrease consistently with increasing SNR, demonstrating that the proposed self-referenced calibration accurately recovers the reference parameters and effectively mitigates the asynchronous offsets.
Fig.~\ref{fig:offset} evaluates the robustness of Group-SBL and JMUSIC by additionally injecting controlled residual TO and common-phase errors into the frontend-calibrated observations obtained in Section~III-A. Fully synchronous perfect baselines are also included for comparison. As the injected residual TO increases, the delay NMSE deteriorates markedly without residual-TO refinement, whereas the proposed alternating residual-TO refinement substantially mitigates this degradation. For Doppler estimation, increasing residual common-phase errors leads to a noticeable degradation without refinement, while the Doppler NMSE remains nearly unchanged for Group-SBL and increases only slightly for JMUSIC when residual common-phase refinement is enabled. The ROR-assisted schemes achieve performance
closer to the fully synchronous baselines, demonstrating
their effectiveness in mitigating residual synchronization errors.

Fig. ~\ref{fig:snr} compares the delay NMSE, Doppler NMSE, and AoA RMSE versus signal-to-noise ratio (SNR). The estimation errors generally decrease as SNR increases. Group-SBL consistently outperforms Individual-SBL, while JMUSIC achieves clear gains over PERMUSIC. These gains result from reinforcing shared-path evidence across UEs and exploiting complementary frequency-domain information from their distinct PRB allocations. ROR further improves both multiuser schemes, with more pronounced gains at high SNRs, where residual calibration errors become more influential as the noise level decreases.  The recorded average global-pool sizes of JMUSIC-GLOBAL at SNRs of \(-4, 0, 4, 8,\) and \(12\) dB are 16.67, 17.59, 17.47, 17.77, and 17.93, consistent with the expected 17 scene-level NLoS components, i.e., \(C_{\rm true}(L_{\rm sh}-1)+KL_{\rm pr}=3\times3+8=17\).

\begin{table}[t]
\centering
\caption{Performance ratios versus the number of allocated subcarriers.}
\label{tab:nk_ratio}
\footnotesize
\renewcommand{\arraystretch}{1.05}
\setlength{\tabcolsep}{2pt}

\begin{tabular}{lrrrrr}
\toprule
Metric &
$N_k=72$ &
$96$ &
$120$ &
$144$ &
$192$ \\
\midrule

\multicolumn{6}{c}{$(L_{\rm sh},L_{\rm pr})=(4,1)$} \\

\multicolumn{6}{l}{Individual-SBL / Group-SBL w/ ROR} \\
Delay NMSE & 2.98 & 3.12 & 3.09 & 3.97 & 4.79 \\
AoA RMSE   & 1.81 & 1.91 & 1.52 & 1.76 & 1.94 \\

\multicolumn{6}{l}{PERMUSIC / JMUSIC w/ ROR} \\
Delay NMSE & 3.91 & 4.53 & 7.45 & 12.40 & 27.00 \\
AoA RMSE   & 1.65 & 2.60 & 2.53 & 2.41 & 3.04 \\

\midrule

\multicolumn{6}{c}{$(L_{\rm sh},L_{\rm pr})=(4,0)$} \\

\multicolumn{6}{l}{Individual-SBL / Group-SBL w/ ROR} \\
Delay NMSE & 6.22 & 12.90 & 24.20 & 20.50 & 37.20 \\
AoA RMSE   & 2.54 & 4.89 & 4.98 & 4.18 & 5.47 \\

\multicolumn{6}{l}{PERMUSIC / JMUSIC w/ ROR} \\
Delay NMSE & 9.35 & 15.50 & 26.70 & 30.70 & 91.20 \\
AoA RMSE   & 3.86 & 5.31 & 6.99 & 8.06 & 10.78 \\

\bottomrule
\end{tabular}
\end{table}

Figs.~\ref{fig:nk41} and \ref{fig:nk40} compare the delay NMSE and AoA RMSE versus the number of allocated subcarriers $N_k$ for $(L_{\rm sh},L_{\rm pr})=(4,1)$ and $(4,0)$, respectively. Both metrics generally decrease with $N_k$ due to enhanced delay-AoA resolution. Under both propagation settings, Group-SBL w/ ROR
and JMUSIC w/ ROR outperform their respective
per-UE baselines. JMUSIC-GLOBAL also improves with increasing $N_k$, particularly in delay estimation.

Table~\ref{tab:nk_ratio} further quantifies these gains. The per-UE baseline-to-multiuser performance ratios are substantially larger for $L_{\rm pr}=0$ than for $L_{\rm pr}=1$, confirming that stronger cross-UE correlation yields greater multiuser gains. For $L_{\rm pr}=1$, the shared-private formulation enables Group-SBL to separately model common and UE-specific propagation components, leading to the best overall delay-AoA estimation performance among the considered schemes. As $L_{\rm pr}$ decreases to zero, both Group-SBL and JMUSIC benefit more from cross-UE evidence aggregation, with the gains becoming more significant as $N_k$ increases and the correlated delay-AoA components are resolved more reliably.

Fig.~\ref{fig:T} compares the delay NMSE and Doppler NMSE versus the
number of CSI snapshots $T$. Both metrics generally decrease as $T$
increases. Group-SBL w/ ROR consistently yields lower delay and Doppler
errors than Individual-SBL, and JMUSIC w/ ROR shows a clear advantage
over PERMUSIC in both metrics. 
JMUSIC-GLOBAL also exhibits decreasing delay NMSE as more snapshots
become available.
Increasing $T$ provides more snapshots for delay-AoA support recovery and extends the coherent slow-time observation interval for Doppler estimation.

Fig.~\ref{fig:K} compares the delay NMSE and AoA RMSE versus the number of UEs $K$ under a fixed total sensing-resource budget ($N=1620$). As $K$ increases, Individual-SBL and PERMUSIC generally deteriorate because fewer subcarriers are available per UE. In contrast, Group-SBL w/ ROR and JMUSIC w/ ROR initially improve from $K=6$ to $K=9$, due to stronger cross-UE evidence, but their errors increase for larger $K$ when the loss of per-UE frequency-domain observations becomes dominant. JMUSIC-GLOBAL exhibits a similar trend, particularly in delay estimation. These results illustrate the tradeoff between cross-UE correlation gain and per-UE sensing-resource reduction under a fixed total budget.

%Fig.~\ref{fig:subcarrier_performance} shows the delay NMSE and AoA RMSE versus the number of allocated subcarriers $N_k$ under$(L_{\mathrm{sh}},L_{\mathrm{pr}})=(4,1)$ and $(4,0)$. Both metrics decrease as $N_k$ increases, and Group-SBL consistently outperforms Individual-SBL under both path configurations. For delay estimation, the performance gap between the two methods is noticeably larger when $L_{\mathrm{pr}}=0$ than when $L_{\mathrm{pr}}=1$. This is because a fully shared support structure allows more paths to benefit from cross-UE evidence aggregation, leading to a larger Group-SBL gain. The results therefore confirm that the benefit of Group-SBL increases with the degree of cross-UE support correlation.

%Fig.~\ref{fig:user_number_performance} compares the delay and Doppler NMSEs versus the number of UEs $K$. To examine the effect of increasing the number of correlated UEs, three propagation groups are fixed, and $K$ is increased from 6 to 15 by adding one UE to each group at every step, with the smaller-$K$ cases retained as nested subsets. As $K$ increases, Group-SBL achieves a steady reduction in both metrics, whereas Individual-SBL and OMP exhibit only minor fluctuations. The additional UEs provide more observations of the group-shared supports, thereby widening the performance gap between Group-SBL and the per-UE baselines.
\section{Conclusion}

This paper developed an asynchronous multiuser uplink sensing framework that exploits cross-UE delay-AoA correlation. Group-SBL explicitly models shared-private sparsity and jointly infers UE grouping, making it suitable for recovering UE-specific propagation structures at the cost of iterative Bayesian inference. In contrast, JMUSIC implicitly exploits cross-UE correlation through joint subspace processing, yielding a global delay-AoA pool that directly supports scene-level sensing without UE-specific association, while enabling UE-specific delay-AoA support recovery when required. However, its performance depends on reliable subspace estimation and global-pool recovery. Both estimators incorporate residual-TO refinement, followed by residual common-phase compensation before Doppler estimation. Simulations demonstrate their advantages over corresponding per-UE baselines, particularly when cross-UE propagation correlation is strong.

\appendices

\section{Proof of Proposition~\ref{prop:joint_music_support}}
\label{app:joint_music_proof}

From
$\boldsymbol{\Pi}_{\mathrm n}
=\mathbf I-\mathbf U_{\mathrm s}\mathbf U_{\mathrm s}^{\mathsf H}$,
$\mathbf C_p=\mathbf U_{\mathrm s}^{\mathsf H}\mathbf B_p$,
$\mathbf G_p=\mathbf C_p^{\mathsf H}\mathbf C_p$, and
$\mathbf B_p^{\mathsf H}\mathbf B_p=\mathbf I_K$, we have
\begin{equation}
\begin{aligned}
\mathbf B_p^{\mathsf H}\boldsymbol{\Pi}_{\mathrm n}\mathbf B_p
&=
\mathbf B_p^{\mathsf H}
(\mathbf I-\mathbf U_{\mathrm s}\mathbf U_{\mathrm s}^{\mathsf H})
\mathbf B_p \\
&=
\mathbf I_K-\mathbf C_p^{\mathsf H}\mathbf C_p
=
\mathbf I_K-\mathbf G_p .
\end{aligned}
\end{equation}
Therefore,
$\lambda_{\min}
(\mathbf B_p^{\mathsf H}\boldsymbol{\Pi}_{\mathrm n}\mathbf B_p)
=
1-\lambda_{\max}(\mathbf G_p)
=
1-\mu_p$,
which yields
$P_{\mathrm J}(p)=1/(1-\mu_p)$.

To characterize how cross-UE evidence contributes to $\mu_p$, we express its largest eigenvalue in the Rayleigh-quotient form
\begin{equation}
\mu_p
=
\max_{\|\mathbf v\|_2=1}
\mathbf v^{\mathsf H}\mathbf G_p\mathbf v
=
\max_{\|\mathbf v\|_2=1}
\sum_{i=1}^{K}\sum_{j=1}^{K}
v_i^{*}v_j[\mathbf G_p]_{i,j},
\end{equation}
where, using
$[\mathbf G_p]_{k,k}=\eta_k(p)$ and
$[\mathbf G_p]_{i,j}
=\mathbf c_{i,p}^{\mathsf H}\mathbf c_{j,p}$ for $i\neq j$,
\begin{equation}
\mathbf v^{\mathsf H}\mathbf G_p\mathbf v
=
\sum_{k=1}^{K}|v_k|^2\eta_k(p)
+
2\operatorname{Re}
\{
\sum_{i<j}
v_i^{*}v_j
\mathbf c_{i,p}^{\mathsf H}\mathbf c_{j,p}
\}.
\end{equation}

The first term combines the UE-specific support values $\eta_k(p)$, whereas the
second captures cross-UE coupling through the off-diagonal
entries of $\mathbf G_p$, which can enhance the joint support
of components observed by multiple UEs.

Since each canonical basis vector $\mathbf e_k$
is feasible in (39), choosing $\mathbf v=\mathbf e_k$ gives $\mu_p
\geq
\mathbf e_k^{\mathsf H}\mathbf G_p\mathbf e_k
=
[\mathbf G_p]_{k,k}
=
\eta_k(p),$
for every $k$. Therefore, $\mu_p\geq\max_k\eta_k(p)$,
which completes the proof.

\vspace{-0.5em}
\setlength{\baselineskip}{12pt}
\bibliographystyle{IEEEtran}
%\bibliography{Ref_ctc,C:/Yu/CA/Papers/Ref}
\bibliography{Reference}
\end{document}